\documentclass[letterpaper,titlepage,11pt]{article}
\pdfoutput=1
\usepackage[colorlinks=true,urlcolor=blue,citecolor=magenta]{hyperref}
\usepackage{amssymb,amsmath,amsfonts}
\usepackage{color}
\usepackage{graphicx}
\usepackage{array}
\usepackage{epsfig}
\usepackage{epstopdf}
\usepackage{caption}
\usepackage{subcaption}
\usepackage{epsfig}
\usepackage{epstopdf}
\usepackage{eso-pic}
\usepackage{tensor}
\usepackage{float}

\def\clock{{\count0=\time
          \divide\count0 60
          \ifnum\count0<10 0\fi\the\count0
          \multiply\count0 -60 \advance\count0 \time
          :\ifnum\count0<10 0\fi \the\count0
        }}
\newcommand{\timestamp}{{\small\vbox{\hbox{\tt\jobname.tex}
\hbox{\the\day/\the\month/\the\year, \clock}}}}

\newcommand{\CB}{\mathcal{B}}

\newcommand{\CD}{\mathcal{D}}

\newcommand{\CL}{\mathcal{L}}

\newcommand{\CT}{\mathcal{T}}
\newcommand{\CR}{\mathcal{R}}
\newcommand{\CS}{\mathcal{S}}

\newcommand{\CP}{\mathcal{P}}

\newcommand{\nn}{\nonumber}

\newcommand{\spa}{\ , \ \ }

\newcommand{\beq}{\begin{equation}}
\newcommand{\eeq}{\end{equation}}

\newcommand{\kk}{{\bf k}}

\numberwithin{equation}{section}

\begin{document}

\begin{titlepage}
\ \ \vskip 1.8cm

\centerline{\LARGE \bf Constraints on the effective fluid theory} 
\vskip 0.3cm
\centerline{\LARGE \bf of stationary branes}

\vskip 1.6cm \centerline{\bf Jay Armas$\,^{1}$ and Troels Harmark$\,^{2}$} \vskip 0.7cm

\begin{center}
\sl $^1$ Albert Einstein Center for Fundamental Physics, University of Bern,\\
Sidlerstrasse 5, 3012 Bern, Switzerland
\vskip 0.3cm
\sl $^2$ The Niels Bohr Institute, Copenhagen University,  \\
\sl  Blegdamsvej 17, DK-2100 Copenhagen \O , Denmark
\end{center}
\vskip 0.3cm

\centerline{\small\tt jay@itp.unibe.ch, harmark@nbi.dk}

\vskip 1.3cm \centerline{\bf Abstract} \vskip 0.2cm \noindent
We develop further the effective fluid theory of stationary branes. This formalism applies to stationary blackfolds as well as to other equilibrium brane systems at finite temperature. The effective theory is described by a Lagrangian containing the information about the elastic dynamics of the brane embedding as well as the hydrodynamics of the effective fluid living on the brane. The Lagrangian is corrected order-by-order in a derivative expansion, where we take into account the dipole moment of the brane which encompasses finite-thickness corrections, including transverse spin.
We describe how to extract the thermodynamics from the Lagrangian and we obtain constraints on the higher-derivative terms with one and two derivatives. These constraints follow by comparing the brane thermodynamics with the conserved currents associated with background Killing vector fields. In particular, we fix uniquely the one- and two-derivative terms describing the coupling of the transverse spin to the background space-time. Finally, we apply our formalism to two blackfold examples, the black tori and charged black rings and compare the latter to a numerically generated solution.

\end{titlepage}

\tableofcontents
\pagestyle{plain}
\setcounter{page}{1}

\section{Introduction}

Physics of branes has been a tremendously useful and important subject in the last two decades. Beyond the fundamental string, there are a number of important types of branes in string and M-theory. At zero temperature the effective description of such branes in the low energy, long distance limit, is highly dependent on which brane one considers, how many there are, and so on. Thus, it is not easy to exhibit universal features of such effective description that are common to all branes. This situation is quite different when one considers branes at finite temperature and in equilibrium, $e.g.$ stationary branes. As has been explored in several recent works, it is possible to find universal features for such branes \cite{Emparan:2009at,Armas:2013hsa, Armas:2013goa}.

One highly important feature of stationary branes at finite temperature is that they can be accurately described by an effective fluid that lives on the embedding surface of the brane. Such fluid branes are an interesting generalisation of general fluids to fluids that are constrained to live on a dynamical surface such that the fluid degrees of freedom can interact with the degrees of freedom of the surface. Fluid branes capture the low energy, long distance dynamics of many interesting and important systems, ranging from blackfolds (black branes wrapped on a dynamical surface) \cite{Emparan:2007wm,Emparan:2009cs,Emparan:2009at} to D-branes at finite temperature (for both weak and strong coupling) \cite{Grignani:2010xm, Emparan:2011hg, Grignani:2013ewa} and biophysical membranes \cite{Armas:2014bia}. The effective theory of fluid branes can also be seen as a generalisation of the fluid/gravity correspondence \cite{Bhattacharyya:2008jc, Emparan:2013ila}.

In this paper we explore the effective fluid description of stationary branes beyond the leading order approximation, following \cite{Armas:2013hsa}. While at leading order the fluid brane is approximated as a surface of zero thickness with a perfect fluid living on the it, at higher orders one needs to take into account corrections to the brane dynamics due to finite-thickness effects related to the elastic nature of the brane, as well as hydrodynamic and transverse spin corrections.
The finite-thickness effects are introduced by expanding the effective stress-energy tensor of the brane in a multipole expansion. We consider here the monopole and dipole contributions and dub the resulting effective description as \emph{pole-dipole fluid branes}. 

Following \cite{Armas:2013hsa}, the effective fluid description of stationary branes is given by a Lagrangian which is corrected order-by-order in a derivative expansion. We focus in this paper on the derivation of constraints for different terms that can appear in the Lagrangian at a given order, in particular, up to and including terms with two derivatives. These constraints are found in part by developing our understanding of how to extract the thermodynamics of the brane from the Lagrangian. Since we consider stationary branes, the system composed of brane plus background has a certain number of Killing vector fields. The conserved currents associated with these Killing vector fields are studied and are an essentially ingredient in the derivation of the constraints on the Lagrangian, as these can be obtained by demanding that the thermodynamics of the brane should match the charges computed using conserved currents. These constraints are not trivial as we are demanding that the long-wavelength dynamics that governs a system with a specific Lagrangian has a pole-dipole expansion. The fact that such constraints exist is a further indication that the effective fluid description of stationary branes exhibits universal features.

Another main focus of this paper is the effective description of branes that are spinning in directions which are transverse to their embedding surface. We develop the effective theory for these cases and use the constraints on the effective Lagrangian to fix the form of the one- and two-derivative terms that describe the transverse spin of the branes. In this way, we find the precise coupling between possible frame-dragging effects on the brane due to the background as well as the additional angular velocity that this causes. Moreover, we find that this coupling completely characterises the one- and two-derivative terms in the effective Lagrangian, $i.e.$ that one can obtain these terms completely by adding an additional angular velocity contribution to the leading order one.

Finally, we apply our effective theory developments to two blackfold examples, namely the black tori originally found in \cite{Emparan:2009cs,Emparan:2009vd} and black rings with Maxwell charge found in \cite{Caldarelli:2010xz}. In both cases, we use the constraints on the effective description to infer the corrected second order thermodynamics of the branes. Both cases can be seen as extensions of the case of neutral black rings studied in \cite{Armas:2014bia}, which was found to be in excellent agreement with the numerical data of \cite{Dias:2014cia}. For the case of black tori we can make this extension due to the fact that the metric on the world-volume is flat, even though the extrinsic curvature is non-zero. For the case of black rings with Maxwell charge, we extend the effective description to include Maxwell charge and compare this to numerical data which we generate by applying an uplift-boost-reduce transformation to the numerical data of the neutral black rings constructed in \cite{Dias:2014cia}. In Sec.~\ref{conclusions} we summarise our results and discuss open directions. We also provide several technical appendices. In App.~\ref{cc} we show the equivalence between space-time and world-volume currents. In App.~\ref{spin} we show how to obtain the spin current conservation from the Lagrangian. In App.~\ref{trans_spin} we present several relations for the derivatives of Killing vector fields. Finally, in App.~\ref{nbr} we provide details on how to construct numerically charged black rings.
Before proceeding, we provide a useful summary of the notation used in the course of this paper in Tab.~\ref{table:notation}.
\\ 
\renewcommand{\arraystretch}{1.5}
\begin{table}[H]
\begin{center}
    \begin{tabular}{ | c | c | c | c | }
    \hline
    \color{red}{Symbol} & \color{red}{Description} &\color{red}{Symbol} & \color{red}{Description}  \\ \hline
    $T^{ab}$ &  world-volume stress-energy tensor & $\textbf{k}^{\mu}$  & background Killing vector\\   \hline
    $\mathcal{D}^{abi}$ & bending moment & $\textbf{k}$ & modulus of the Killing vector  \\   \hline
    $\mathcal{S}^{aij}$ & spin current  & ${u_a}^{\mu}$ & parallel projector onto the brane \\   \hline 
    $\gamma_{ab}$ & induced metric  & ${n_{i}}^{\mu}$ & orthogonal projector to the brane \\   \hline
    ${K_{ab}}^{i}$ & extrinsic curvature tensor & $\xi^{\mu}$ & time-like Killing vector \\   \hline
    ${\omega_{a}}^{ij}$ & extrinsic twist potential & $\chi^{\mu}_q$ & rotational world-volume Killing vector \\   \hline
     $\mathcal{R}_{ab}$ & world-volume Ricci tensor & $\hat\chi^{\mu}_{\hat q}$ & rotational transverse Killing vector  \\   \hline
     $\mathcal{R}$ & world-volume Ricci scalar  & $\Omega_q$ & angular velocity associated with $\chi^{\mu}_q$  \\   \hline
     ${\Omega_{ab}}^{ij}$ & outer curvature of the embedding & $\hat\Omega_{\hat q}$ & angular velocity associated with $\hat\chi^{\mu}_{\hat q}$ \\   \hline
     \end{tabular}
     \caption {Summary of the notation used in this paper.}  \label{table:notation}
\end{center} 
\end{table}
\vskip 0.3cm

\section{Thermodynamics of stationary fluid branes}
\label{sec:thermo}

A fluid brane is a long-wavelength effective description of the physics of a brane in a background space-time. This effective description can describe a blackfold ($i.e.$ a black brane wrapped on a dynamical surface), D-branes at finite temperature ($i.e.$ thermal DBI) \cite{Grignani:2010xm, Emparan:2011hg, Grignani:2013ewa} or any of the other type of branes in string/M-theory as well as biophysical membranes \cite{Armas:2014bia} or cosmic strings \cite{Carter:1997pb, Anderson:1997ip}.

In the following, after a brief digression on Killing vector fields of stationary fluid branes, we first review the thermodynamics of stationary fluid branes in the perfect fluid limit, and subsequently postulate a generalisation to stationary fluid branes with pole-dipole corrections including up to two-derivative terms. 

\subsection{Thermodynamics of pole-dipole fluid branes} \label{thermo}


Consider a space-time with metric $g_{\mu\nu}$, where the greek indices $\mu,\nu...$ label space-time directions. We consider a fluid $p$-brane with embedding $X^\mu(\sigma)$ where $\sigma^a$, $a=0,1,...,p$, are the world-volume coordinates. The induced metric on the world-volume is $\gamma_{ab}=\partial_a X^\mu \partial_b X^\nu g_{\mu\nu}$ and the fluid velocity on the brane is denoted by $u^a$ such that $\gamma_{ab} u^a u^b =-1$.

A stationary fluid brane is characterized by a Killing vector field $\kk^\mu$ of the background space-time. 
The space-time Killing vector field $\kk^\mu$ is mapped to the world-volume vector $\kk^a = \gamma^{ab} g_{\mu\nu} \partial_b X^\mu \kk^\nu$. For the fluid brane to be stationary one requires that $\kk^a$ is a Killing vector field on the world-volume with respect to the induced metric $\gamma_{ab}$ and that the fluid velocity $u^a$ is proportional to $\kk^a$ thus ensuring the absence of dissipation. Specifically, this means $u^a = \kk^a / {\bf k}$ where ${\bf k} = \sqrt{-\gamma_{ab} \kk^a \kk^b}$. Thus, by \emph{stationary fluid brane} we mean that the total system of brane plus background space-time possesses $\kk^\mu$ as a Killing vector field, including that the induced vector $\kk^a$ is a Killing vector field on the world-volume and that the fluid velocity $u^a$ is proportional to it.

We consider a space-time with a time-translation Killing vector field $\xi$ and a number of rotational Killing vector fields (spatial Killing vector fields with closed orbits and a submanifold of fixed points). 
According to the embedding $X^\mu(\sigma)$ the rotational Killing vector fields can be divided into those that have a non-trivial orbit on the world-volume (Killing vector fields denoted by $\chi_q$) and those for which the embedding is contained in the submanifold of fixed points (Killing vector fields denoted by $\hat{\chi}_{\hat{q}}$). Here $q$ and $\hat{q}$ are indices labelling the set of commuting and linearly independent rotational Killing vector fields. We assume that the space-time Killing vector field $\kk^\mu$ is a linear combination of these Killing vector fields
\begin{equation}
\label{KVF_bulk}
\kk^\mu = \xi^\mu + \sum_q \Omega_q \chi_q^\mu + \sum_{\hat{q}} \hat{\Omega}_{\hat{q}} \hat{\chi}_{\hat{q}}^\mu~~,
\end{equation}
where $\Omega_q$ and $\hat{\Omega}_{\hat{q}}$ are constants corresponding to the angular velocities of the fluid brane.
Mapping $\kk^\mu$ to the world-volume we find
\begin{equation}
\label{KVF_wv}
\kk^a = \xi^a + \sum_q \Omega_q \chi_q^a~~.
\end{equation}
The part with $\hat{\chi}_{\hat{q}}^\mu$ is not there since the map to the world-volume gives a vector with zero norm. We assume that $\xi^a$ and $\chi_q^a$ are Killing vector fields on the world-volume which then guarantees that $\kk^a$ is a Killing vector field on the world-volume. Defining the norms
\begin{equation}
\gamma_{ab} \xi^a \xi^b = - R_0^2 \spa \gamma_{ab} \chi^a_q \chi^b_q = R_q^2~~,
\end{equation}
we find
\begin{equation}
{\bf k} = \sqrt{R_0^2 - \sum_q \Omega_q^2 R_q^2}~~.
\end{equation}
%

\subsubsection*{More on geometry of brane embeddings}

First we briefly go through the various geometric objects associated with brane embeddings that we need in this paper, in addition to those already mentioned above.
We introduce the notation $u_a {}^\mu = \partial_a X^\mu$ for vectors longitudinal to the brane and the projector $\gamma^{\mu\nu} = u_a {}^\mu u_b {}^\nu \gamma^{ab}$.
We define the transverse normalized vectors $n_i {}^\mu$ such that $n_i {}^\mu u_{a} {}^\nu g_{\mu\nu} = 0$ and $n_i {}^\mu n_j {}^\nu g_{\mu \nu} = \delta_{ij}$ as well as the transverse projector $\perp_{\mu\nu} = g_{\mu\nu} - \gamma_{\mu\nu} = n^i {}_\mu n^j {}_\nu \delta_{ij}$. Here the latin indices $i,j,...$ label transverse directions to the brane embedding. Note that for any vector we can write $v^\mu = u_a {}^\mu v^a + n_i {}^\mu v^i$ and similarly we can split up any other index of a tensor in longitudinal and transverse parts. In particular, we use the notation $\nabla_\mu = u^a {}_\mu \nabla_a + n^i {}_\mu \nabla_i$ for the background covariant derivative (note that $\nabla_a$ is not the covariant derivative on the world-volume). Furthermore, the extrinsic curvature tensor is $K_{\mu\nu} {}^\rho = \gamma_\mu {}^\lambda \gamma_\nu {}^\sigma \nabla_ \lambda \gamma_\sigma {}^\rho = K_{ab} {}^i u^a {}_\mu u^b {}_\nu n_i {}^\rho$ and the extrinsic twist potential is $\omega_a {}^{ij} = n^i {}_\mu \nabla_a n^{j \mu}$ which is antisymmetric in the $i$ and $j$ indices.

Each of the Killing vector fields $\hat{\chi}_{\hat{q}}$ corresponds to a certain rotation plane of the background geometry (in an appropriately chosen coordinate system). We call these planes the {\sl transverse spin planes} of the brane. The fact that the brane embedding is at a fixed point of $\hat{\chi}_{\hat{q}}$ means that it is situated at the centers of all the rotation planes (which makes it clear why we denote the planes as transverse). In other words, the brane embedding does not bend into transverse spin planes. 

Another way to see that the brane embedding does not bend into the transverse spin planes is as follows. Since $\hat{\chi}_{\hat{q}}$ is transverse to the brane one has $\gamma_\mu {}^\nu \hat{\chi}_{\hat{q},\nu}=0$. Taking the derivative this gives $\gamma_\lambda {}^\rho \gamma_\sigma {}^\mu \nabla_\rho (\gamma_\mu {}^\nu \hat{\chi}_{\hat{q},\nu})=0$ thus giving $K_{\lambda\sigma} {}^\nu \hat{\chi}_{\hat{q},\nu} = - \gamma_\lambda {}^\rho \gamma_\sigma {}^\nu \nabla_\rho \hat{\chi}_{\hat{q},\nu}$. Using that the extrinsic curvature tensor is symmetric in the longitudinal indices and that $\hat{\chi}_{\hat{q}}$ is a Killing vector field one infers that $K_{ab} {}^\rho \hat{\chi}_{\hat{q},\rho} = 0$, from which we obtain
\begin{equation}
\label{nonbend1}
K_{ab} {}^i \hat{\chi}_{\hat{q},i} = 0~~.
\end{equation}

The physical meaning of the extrinsic twist potential is that it parameterizes the possible frame-dragging that the brane is subject to in each of the transverse spin planes due to the background. One can choose a frame for the transverse vectors $n_i {}^\mu$ such that each transverse spin plane corresponds to two specific values of $i$. The extrinsic twist potential then picks up contributions from all the transverse spin planes. Define for each transverse spin plane the Levi-Civita symbol $\epsilon_{ij}^{(\hat{q})}=\epsilon^{ij}_{(\hat{q})}$ such that it is only non-zero ($e.g.$ plus or minus one) if $i,j$ runs over the two specific values corresponding to the transverse spin plane and where the index $\hat{q}$ means that it is the rotation plane corresponding to the $\hat{\chi}_{\hat{q}}$ rotational Killing vector field. Then we write the extrinsic twist potential for the brane embedding as
\begin{equation}
\omega_a {}^{ij} = \sum_{\hat{q}} \omega^{(\hat{q})}_a \epsilon^{ij}_{(\hat{q})}~~,~~\omega^{(\hat{q})}_a=\frac{1}{2}\epsilon_{ij}^{(\hat{q})}\omega_a {}^{ij}~~,
\end{equation}
where $\omega^{(\hat{q})}_a$ is the normal fundamental form associated with the transverse spin plane with Killing vector field $\hat{\chi}_{\hat{q}}$.

\subsubsection*{Action for stationary pole-dipole fluid brane}

In this paper we consider the thermodynamics of stationary fluid branes beyond the perfect fluid limit by including the so-called pole-dipole terms.
Going beyond the perfect fluid limit of a fluid brane various new effects will play a role in the dynamics. A very efficient and powerful method to describe these effects for stationary fluid branes is by using the action principle \cite{Armas:2013hsa}. One reason for corrections to the perfect fluid dynamics is due to finite-thickness corrections of the fluid brane. This is reminiscent of the multipole expansion for post-Newtonian physics from General Relativity. The leading part is the monopole contribution corresponding to an infinitely thin brane. The first correction comes from the dipole contribution which one parameterizes by the dipole moment $\CD^{ab}{}_i$ and the spin current $\CS^a{}_{ij}$. For simplicity, we do not consider multipoles beyond the dipole. The monopole moment $T^{ab}$ as well as the higher moments $\CD^{ab}{}_i$ and $\CS^a{}_{ij}$ can be obtained from the action when one makes variations with respect to certain geometric quantities. Indeed, one finds \cite{Armas:2013hsa}
\begin{equation} \label{Tab}
T^{ab} = \frac{2}{\sqrt{-\gamma}} \frac{\partial (\sqrt{-\gamma} \CL)}{\partial \gamma_{ab}} \spa 
\CD^{ab}{}_i = \frac{\partial \CL}{\partial K_{ab}{}^i} \spa \CS^a{}_{ij} = \frac{\partial \CL}{\partial \omega_{a}{}^{ij}}~~,
\end{equation} 
where the action is $I = - \beta \int_{\CB_p} dV_{(p)} R_0 \CL$. When including the dipole correction to the stress-energy distribution of the fluid brane the action thus has the following schematic form
\begin{equation}
\label{fullaction}
I [T, \kk^a , \hat{\Omega}_{\hat{q}}    ,  \gamma_{ab}, K_{ab}{}^i, \omega_a{}^{ij}] =  \beta \int dV_{(p)} R_0 \CL ( T, \kk^a, \hat{\Omega}_{\hat{q}}    ,  \gamma_{ab}, K_{ab}{}^i, \omega_a{}^{ij}, \nabla^a)~~,
\end{equation}
with $T$ being the global temperature and 
where $\nabla^a$ in the Lagrange function $\CL$ is a formal way of writing that the Lagrangian also depends on the world-volume derivatives of the quantities, $e.g.$ it can for instance depend on $\nabla^a \kk^b$.
Up to two derivatives along the world-volume we can write the stationary fluid brane Lagrangian as%
\footnote{Note that a possible term proportional to $u_a \omega^{a ij} u^b \omega_{b ij}$, as studied in \cite{Armas:2013hsa}, is a particular case of the term proportional to $\varpi_2^{(\hat{q},\hat{q}')}$ in the Lagrangian \eqref{lagrangian}. Furthermore, in writing the possible terms in \eqref{lagrangian} we have assumed that the surface codimension is higher than one and we have neglected higher-pole moments beyond dipole contributions, in particular this implies that we ignore terms involving $\nabla_b\omega^{(\hat{q})}_a$. Some of the possible contributions involving $\omega^{a ij}$ can be eliminated by using the constraints that will be derived in Sec.~\ref{sec:constraints} while others can be \emph{a priori} removed as they do not satisfy the conservation equation \eqref{spinconservation} for the spin current.} 
\begin{eqnarray}
\label{lagrangian}
\CL &=& \lambda_0 + \upsilon_1 \omega^{ab} \omega_{ab} + \upsilon_2 \CR + \upsilon_3 u^a u^b \CR_{ab} + \lambda_1 K^i K_i + \lambda_2 K^{abi} K_{abi} + \lambda_3 u^a u^b K_a{}^{ci}K_{bci} \nn \\ && + \sum_{\hat{q}} \varpi^{(\hat{q})}_1  u^a \omega^{(\hat{q})}_a +  \sum_{\hat{q},\hat{q}'} \varpi_2^{(\hat{q},\hat{q}')} u^a \omega^{(\hat{q})}_a u^b \omega^{(\hat{q}')}_b ~~,
\end{eqnarray}
where $\omega_{ab}$ is the vorticity of the fluid velocity $u^a$ and $\mathcal{R}_{ab}$ and $\mathcal{R}$ are the Ricci tensor and Ricci scalar associated to the world-volume metric.\footnote{One should also consider in \eqref{lagrangian} possible couplings to the background Riemann curvature tensor. However, such couplings can be exchanged by couplings to the worldvolume Riemann curvature tensor and couplings to the extrinsic curvature tensor via Gauss-Codazzi equations. This has been analysed in detail in \cite{Armas:2013hsa}.} To each of the higher-order contributions beyond $\lambda_0$ we associate the parameter $\tilde\varepsilon$ which counts the number of derivates, e.g., the term $\varpi^{(\hat{q})}_1  u^a \omega^{(\hat{q})}_a$ is the only term in \eqref{lagrangian} of order $\mathcal{O}(\tilde\varepsilon)$.
The Gibbs free energy is then
\begin{equation}
\label{gibbsfreeenergy}
F [T, \kk^a , \hat{\Omega}_{\hat{q}}    ,  \gamma_{ab}, K_{ab}{}^i, \omega_a{}^{ij}] = -  \int dV_{(p)} R_0 \CL ( T , \kk^a, \hat{\Omega}_{\hat{q}}    ,  \gamma_{ab}, K_{ab}{}^i, \omega_a{}^{ij}, \nabla^a)~~.
\end{equation}
%

\subsubsection*{Thermodynamics}

We consider now how to extract the global thermodynamics of the stationary fluid brane given the action \eqref{fullaction} with Lagrange function \eqref{lagrangian}, or, equivalently, with free energy \eqref{gibbsfreeenergy}. Since we are describing a fluid brane there exists a fluid rest-frame, where we are comoving with the fluid velocity $u^a$. In this rest-frame we can measure a local temperature $\CT$ and a local angular velocity $\Omega'_{\hat{q}}$ which are related to the global temperature $T$ and angular velocity $\hat{\Omega}_{\hat{q}}$ via
\begin{equation}
\label{localTandOm}
\CT = \frac{T}{{\bf k}}
\spa \Omega'_{\hat{q}} = \frac{\hat{\Omega}_{\hat{q}}}{{\bf k}}~~.
\end{equation}
The ${\bf k}$ factor appearing in these formulas can be seen as a combined boost and redshift factor due to the choice of being in the fluid rest-frame. The Lagrange function \eqref{lagrangian} depends on the nine scalar functions $\lambda_0$, $\lambda_i$, $\upsilon_i$, $i=1,2,3$ and $\varpi_i$, $i=1,2$. These scalar functions should transform as scalars on the world-volume. As a matter of definition they cannot depend on the tensor quantities used above that transform non-trivially on the world-volume (since \eqref{lagrangian} precisely is an expansion in terms of those quantities). Hence, they can only depend on the local quantities $\CT$ and $\Omega'_{\hat{q}}$ (since a scalar quantity it should be the same whether we are in the rest frame or not).
Using this requirement, as well as the relation $\kk^a (\partial u^b / \partial \kk^a )= 0$, one easily derives the following general identity for the Gibbs free energy \eqref{gibbsfreeenergy}
\begin{equation}
\label{F_relation}
\left[ \kk^a \frac{\partial}{\partial{\kk^a}} + T \frac{\partial}{\partial T} + \sum_{\hat{q}} \hat{\Omega}_{\hat{q}} \frac{\partial}{\partial {\hat{\Omega}_{\hat{q}}}} \right] F =0~~.
\end{equation}
Thus, this identity is a direct consequence of describing the fluid on a brane in the rest-frame.
Employing this identity one can derive the thermodynamic quantities from the Gibbs free energy \eqref{gibbsfreeenergy} as follows
\begin{equation}
\label{thermodyn}
M = F + \xi^a \frac{\partial F}{\partial \kk^a}
\spa S= - \frac{\partial F}{\partial T} = \frac{1}{T} \left[ \kk^a \frac{\partial F}{\partial \kk^a} + \sum_{\hat{q}} \hat{\Omega}_{\hat{q}} \frac{\partial F}{\partial {\hat{\Omega}_{\hat{q}}}} \right]  \spa J_q = - \chi_q^a \frac{\partial F}{\partial \kk^a} \spa \hat{J}_{\hat{q}} = - \frac{\partial F}{\partial {\hat{\Omega}_{\hat{q}}}}~~,
\end{equation}
where $M$ and $S$ are the total mass and entropy and $J_q$ and $\hat{J}_{\hat{q}}$ are the angular momenta due to the  longitudinal and transverse motion of the fluid living on the brane, respectively. With this, the first law in terms of the Gibbs free energy holds
\begin{equation}
\label{totalGibbs1}
dF = - S dT - \sum_q J_q d\Omega_q - \sum_{\hat{q}} \hat{J}_{\hat{q}} d\hat{\Omega}_{\hat{q}}~~,
\end{equation}
when keeping fixed the geometric quantities $\gamma_{ab}$, ${K_{ab}}^i$ and ${\omega_a}^{ij}$ under the variation.

\subsection{Equivalence with conserved currents} \label{sec:currents}
The formulas presented in the previous section allow us to obtain expressions for the thermodynamic properties of generic stationary configurations. On the other hand, given the world-volume theory, obtained by variation of the free energy \eqref{gibbsfreeenergy}, it is possible to obtain a set of conserved surface currents via appropriate contractions with the mutually independent Killing vector fields of the background \cite{Armas:2013hsa}. This world-volume theory was shown to be derived from a space-time formulation, to pole-dipole order, of the stress-energy tensor \cite{Armas:2013hsa}. Therefore, it is relevant to show that, for consistency, conserved space-time currents can be identified with world-volume surface currents, a fact which is demonstrated in App.~\ref{cc}. Given this identification, we show in this section that this set of world-volume currents yields, in absence of transverse spin, the same result for the conserved charges as that obtained from the formulas derived in the previous section. The case of fluid branes with transverse spin is analysed in Sec.~\ref{sec:constraints}. We note that this comparison is \emph{a priori} non-trivial as we are demanding that the existence of certain global potentials such as $\hat\Omega_{\hat q}$ charactering the stationary fluid brane configuration is due to the existence of certain pole-dipole currents such as $\mathcal{S}^{aij}$.

\subsubsection*{Conserved currents of pole-dipole fluid branes}

To each independent space-time Killing vector field $k^{\mu}$ one can associate a surface current $\CP^\nu_{k}$ given by the general expression \cite{Armas:2013hsa}
\begin{equation} \label{conc}
\CP^\nu_{k} = ( T^{ab} u_a^\mu u_b^\nu + \CD^{ab i} {K^{\mu}}_{bi}{u^{\nu}}_{a}+2{u^{\nu}}_{b}\CS^{c\mu i}{K^{b}}_{ci}  -{\mu^{\nu}}_{b}{\perp^{\mu}}_{\lambda}\nabla_a\CD^{ab\lambda}) k_\mu - \CS^{c \mu \rho} {u^\nu}_{c} \nabla_\mu k_\rho + \CD^{\nu\mu\rho} \nabla_{\mu} k_\rho~~,
\end{equation}
which satisfies the world-volume conservation equation ${\gamma^{\mu}}_{\lambda}\nabla_{\mu}\CP^{\nu}_{k}=0$. Given the set of conserved currents \eqref{conc} associated with an arbitrary Killing vector field $k^{\mu}$ one can evaluate the corresponding conserved charge via
\beq \label{ccharge}
|\mathcal{Q}_{k}|=\int_{\mathcal{B}_p}dV_{(p)}\CP^{\nu}_{k}n_{\nu}~~,
\eeq
where $n_{\nu}$ is the unit normalised timelike Killing vector field $n_\nu=\xi_\nu/R_0$. By choosing $k^{\mu}=\xi^{\mu}$ one obtains the mass $M$, while by choosing $k^\mu = \chi_q^\mu$ one obtains the longitudinal angular momenta $J_q$ and choosing $k^\mu = \hat{\chi}_{\hat{q}}^\mu$ one obtains the transverse angular momenta $\hat{J}_{\hat{q}}$.
For $k^{\mu}=\xi^{\mu}$ and $k^\mu = \chi_q^\mu$ one finds the conserved currents
\begin{equation}
\label{long_cur}
\begin{array}{c}
\CP^\nu_{\xi} = T^{ab}u_b {}^\nu\xi_a + 2 u_{a} {}^{\nu} \CD^{ab i} {K^{c}}_{bi}\xi_{c} + u_a {}^\nu \CS^{aij} \omega_{bij} \xi^b ~~, \\[2mm]  \CP^\nu_{\chi_q} = T^{ab}u_b {}^\nu\chi_{q,a} + 2 u_{a} {}^{\nu} \CD^{ab i} {K^{c}}_{bi}\chi_{q,c} +  u_a {}^\nu \CS^{aij} \omega_{bij} \chi_q^b ~~,
\end{array}
\end{equation}
since these Killing vector fields are longitudinal to the brane. The last term in both currents is derived from the fifth term of \eqref{conc} using the identities \eqref{nablaxi} derived in Appendix \ref{trans_spin}.

One can also simplify the conserved currents for $k^\mu = \hat{\chi}_{\hat{q}}$. This simplification follows from the fact that the brane embedding does not bend into the transverse spin planes as explained in Sec.~\ref{thermo}.
This gives Eq.~\eqref{nonbend1} as well as
\beq
\label{nonbend2}
\CD^{abi} \hat{\chi}_{\hat{q},i}=0~~,~~\CD^{abi}\omega_{cij}=0~~,~~\mathcal{S}^{aij}{K_{bcj}}=0~~,
\eeq
where the first identity follows from \eqref{nonbend1} using \eqref{lagrangian} and \eqref{Tab} which shows that $\CD^{abi} \propto {L^{ab}}_{cd}K^{cdi}$ for some tensor ${L^{ab}}_{cd}$. The second and third identities follow from noting that the transverse indices of $\omega_{cij}$ and $\mathcal{S}^{aij}$ are supported on the transverse spin planes. Using \eqref{nonbend1} and \eqref{nonbend2} we find
\begin{equation}
\label{trans_cur}
\CP^\nu_{\hat{\chi}_{\hat{q}}} = - \CS^{c \mu \rho} {u^\nu}_{c} \nabla_\mu \hat{\chi}_{\hat{q},\rho}~~.
\end{equation}

\subsubsection*{Computing $TS$ in terms of the conserved currents}

By choosing $k^{\mu}=\bf k^{\mu}$ we can obtain an equivalent formula to \eqref{F_relation}. In order to do so we split the stress-energy tensor $T^{ab}$ into a perfect fluid part $\mathcal{T}^{ab}$ and a correction $\Pi^{ab}$, as in \cite{Armas:2013hsa}, such that $T^{ab}=\mathcal{T}^{ab}+\Pi^{ab}$. For convenience we introduce the quantities $\mathcal{P}$ and $\mathcal{E}$ such that
\beq \label{stpf}
\mathcal{T}^{ab}=\mathcal{P}\gamma^{ab}+(\mathcal{E}+\mathcal{P})u^{a}u^{b}~~.
\eeq
Since the stress-energy tensor $T^{ab}$ can be obtained from the action $I$ using formula \eqref{Tab} it is easy to see that $\mathcal{P}=\mathcal{L}$. 
Now we proceed and evaluate the r.h.s. of \eqref{ccharge} using the the definition of mass and angular momenta with $k^{\mu}=\bf k^{\mu}$ in order to find
\beq \label{pm}
\int_{\mathcal{B}_p}dV_{(p)}\CP^{\nu}_{\bf k}n_{\nu}=M-\sum_{q}\Omega_q J_q-\sum_{\hat q}\hat\Omega_{\hat q}\hat J_{\hat q}~~.
\eeq
Using the explicit forms of the currents given in Eqs.~\eqref{long_cur} and \eqref{trans_cur}, along with \eqref{stpf}, we obtain an expression for the product $TS$ in the form \footnote{Note here that we are assuming that the stress-energy tensor is obtained from the action via the formula \eqref{Tab} and hence it does not come in the Landau frame, i.e. $\Pi^{ab}u_{b}\ne0$. For details on frame transformations see \cite{Armas:2013goa}.}
\beq \label{TSc}
TS=-\int_{\mathcal{B}_p}dV_{(p)}\left(\textbf{k}(\mathcal{E}+\mathcal{P})u^{b}-\textbf{k}\Pi^{ab}u_{a}-2\CD^{bci}{K^{a}}_{ci}\textbf{k}_a- \CS^{bij} \omega_{aij}\textbf{k}^a+\sum_{\hat q}\CS^{bij}\partial_{i}\hat{\chi}_{\hat{q},j}\right)n_{b}~~,
\eeq
where we have used the Killing equation $\nabla_{\mu}\textbf{k}_{\nu}=\nabla_{[\mu}\textbf{k}_{\nu]}$ in order to exchange covariant derivatives by partial derivatives. This expression is the equivalent to \eqref{F_relation} but written in terms of  conserved surface charges. In order to show this exilictely, we have taken each of the contributions to the free energy \eqref{gibbsfreeenergy} one-by-one and computed the stress-energy tensor, bending moment and spin current using \eqref{Tab}. This has been done explicitly in \cite{Armas:2013hsa} for the terms in \eqref{gibbsfreeenergy} involving $\lambda_0,\upsilon_i,\lambda_i,$, $i=1,2,3$. In Tab. \ref{table:stress}, we present these quantities for the terms involving $\varpi_1, \varpi_2$.
\\ 
\renewcommand{\arraystretch}{1.5}
\begin{table}[H]
\begin{center} 
    \begin{tabular}{ | c | c | c |}
    \hline
    \color{red}{Scalar} & \color{red}{$ T^{ab}$} & \color{red}{${\mathcal{S}^{a}}_{ij}$} \\ \hline
    $\varpi^{(\hat{q})}_1  u^a \omega^{(\hat{q})}_a$ &  $\left(\varpi_1^{(\hat q)}\gamma^{ab}-(\textbf{k}\varpi_1^{(\hat{q})'}-\varpi_1^{(\hat q)})u^{a}u^{b}\right)u^{c}\omega_c^{(\hat q)}$ & $\varpi_1^{(\hat q)} u^{a}\epsilon^{(\hat q)}_{ij}$ \\   \hline
    $\!\!\varpi_2^{(\hat{q},\hat{q}')} u^a \omega^{(\hat{q})}_a u^b \omega^{(\hat{q}')}_b\!\!$ & $\!\!\!\left(\varpi_2^{(\hat{q},\hat{q}')}\gamma^{ab}-(\textbf{k}\varpi_2^{(\hat{q},\hat{q}')'}-2\varpi_2^{(\hat{q},\hat{q}')})u^{a}u^{b}\right)u^c \omega^{(\hat{q})}_c u^d \omega^{(\hat{q}')}_d\!\!\!$ & $\!\!\varpi_2^{(\hat{q}',\hat{q}')}u^{a}u^{c}(\omega_{c}^{(\hat q)}\epsilon_{ij}^{(\hat q)}+\omega_{c}^{(\hat q)}\epsilon_{ij}^{(\hat q')})\!\!$ \\   \hline
     \end{tabular}
     \caption{World-volume stress-energy tensor and spin current obtained from the Lagrangian \eqref{lagrangian} due to the contributions of the scalars involving the transport coefficients $\varpi^{(\hat{q})}_1$ and $\varpi_2^{(\hat{q},\hat{q}')}$. Here the prime in $\varpi_1^{(\hat{q})'},\varpi_2^{(\hat{q},\hat{q}')'}$ means that we have taken a derivative with respect to $\textbf{k}$. } \label{table:stress}
\end{center} 
\end{table}
For these contributions, the bending moment $\mathcal{D}^{abi}$ vanishes by definition. One can also show that $T^{ab}$ and ${\mathcal{S}^{a}}_{ij}$ satisfy the non-trivial identities
\beq \label{spinconservation}
\nabla_aT^{ab}=-{\mathcal{S}^{a}}_{ij}{\Omega_a}^{bij}~~,~{n_\mu}^{i}{n_\nu}^{j}\nabla_a\mathcal{S}^{a\mu\nu}=0~~,
\eeq
where ${\Omega_a}^{bij}$ is the outer curvature of the embedding defined in App.~\ref{spin}. The first equation above follows from invariance under world-volume reparameterizations of the action \eqref{fullaction} as shown in \cite{Armas:2013hsa}, while the second equation follows from requiring the action to be invariant under different choices of the normal vectors. We show that this is indeed the case in App.~\ref{spin}. 

Using the above table together with \eqref{stpf} and introducing it in the r.h.s. of \eqref{TSc} for each term appearing in \eqref{gibbsfreeenergy} leads to the same result obtained using the formula \eqref{F_relation}. This agreement is also verified for the charges computed from \eqref{thermodyn} and those obtained from \eqref{ccharge}. The case of charges associated with $\varpi_1, \varpi_2$, this agreement is only verified provided we constrain the transport coefficients in an appropriate way. These cases will be analysed in detail in Sec.~\ref{sec:constraints}.

\subsubsection*{The entropy current}

From the expression for the product $TS$ given in \eqref{TSc}, we can obtain an explicit formula for the entropy current $J^{a}_{s}$ of stationary fluid branes
\beq \label{scurrent}
J^{a}_{s}=\frac{1}{T}\left(\textbf{k}(\mathcal{E}+\mathcal{P})u^{a}-\textbf{k}\Pi^{ab}u_{b}-2\CD^{aci}{K^{b}}_{ci}\textbf{k}_b- \CS^{aij} \omega_{bij}\textbf{k}^b+\sum_{\hat q}\CS^{aij}\partial_{i}\chi_{j}^{\hat q}\right)~~,
\eeq
from which the total entropy $S$ can be evaluated via the formula
\beq \label{Sf}
S=-\int_{\mathcal{B}_p}dV_{(p)}J^{a}_sn_a~~.
\eeq
The entropy current of stationary configurations is by definition a conserved current. To see explicitly that this is the case we rewrite the entropy current using \eqref{pm}, \eqref{long_cur} and \eqref{trans_cur} as 
\beq
J^{a}_{s}=\frac{1}{T}\left(\textbf{k}\mathcal{P}u^{a}-\mathcal{P}^{a}_{\textbf{k}}\right)~~.
\eeq
It is now clear that this current is conserved, a fact that follows trivially from stationarity of the configuration and the conservation law for $\mathcal{P}^{a}_{\textbf{k}}$.

\subsection{Adding charge} \label{charge}
In this section we briefly generalise the results of the previous sections to the cases of stationary fluid branes carrying either $q=0$-brane charge or $q=p$-brane charge. This type of fluids was studied, for example, in \cite{Grignani:2010xm, Caldarelli:2010xz, Emparan:2011hg}. Other types of charged fluids can be encompassed within the current framework but we leave the details for future work. 

Fluids carrying these types of charges are characterised by a local chemical potential $\Phi_{q}$ and a charge density $\mathcal{Q}_p$ in the case $q=0$ or a global charge $Q_{p}$ in the case $p=q$. This global charge can be obtained by integrating conserved charge currents $J^{a_1...q_{p+1}}_{q}$ over the world-volume. For the case $q=0$, for example, this can be obtained by computing the r.h.s. of \eqref{Sf} with $J^{a}_s$ replaced by $J^{a}_{q}$. From the local potentials $\Phi_q$ one can define a global chemical potential $\Phi_{H}$ via
\beq
\Phi_{H}=\frac{\Phi_0}{\textbf{k}}~~(\text{if} ~q=0)~~,~~\Phi_{H}=\int_{\mathcal{B}_p}dV_{(p)}R_0 \Phi_{q}~~(\text{if} ~q=p)~~.
\eeq

In order to generalise the free energy \eqref{gibbsfreeenergy} to these cases it is just necessary to consider the scalar functions $\lambda_0$, $\lambda_i$, $\upsilon_i$, $i=1,2,3$ and $\varpi_i$, $i=1,2$, to depend now on the local quantities $\CT$, $\hat{\Omega}_{\hat{q}}$ and $\Phi_q$. With this in hand we can find a generalisation of \eqref{F_relation} such that
\begin{equation}
\label{F_relationq}
\left[ \kk^a \frac{\partial}{\partial{\kk^a}} + T \frac{\partial}{\partial T} + \sum_{\hat{q}} \hat{\Omega}_{\hat{q}} \frac{\partial}{\partial {\hat{\Omega}_{\hat{q}}}} +\Phi_{H}\frac{\partial }{\partial \Phi_H}\right] F =0~~.
\end{equation}
Due to the presence of a conserved global charge and the existence of a global chemical potential, the first law of thermodynamics is now
\begin{equation} \label{tq}
dF = - S dT - \sum_q J_q d\Omega_q - \sum_{\hat{q}} \hat{J}_{\hat{q}} d\hat{\Omega}_{\hat{q}}-Q_p d\Phi_H~~,~~Q_p=-\frac{\partial F}{\partial \Phi_H}~~.
\end{equation}
Eq.~\eqref{F_relationq} has also a counterpart in terms of conserved surface currents, in fact, the r.h.s. of Eq.~\eqref{TSc} is left unmodified while the l.h.s. becomes a sum of the contributions $TS+\Phi_H Q_p$. This implies that the entropy current is no longer given by \eqref{scurrent}, instead we use \eqref{thermodyn} and \eqref{tq} in order to write the generic entropy and charge currents as
\beq
J^{a}_s=\textbf{k}\frac{\partial \mathcal{L}}{\partial T}u^{a}~~,~~J^{a}_q=\textbf{k}\frac{\partial \mathcal{L}}{\partial \Phi_H}u^{a}~~(\text{if}~q=0)~~,~~J^{a_1...a_{p+1}}_q=\frac{\partial \mathcal{L}}{\partial \Phi_H}\epsilon^{a_1...a_{p+1}}~~(\text{if}~q=p)~~,
\eeq
which are conserved due to the stationarity of the overall configuration.

\subsubsection*{The Smarr relation}
Since fluid branes can describe the dynamics of black holes, it is interesting to derive an expression for the Smarr relation of these configurations.  Using \eqref{gibbsfreeenergy} and \eqref{ccharge} we obtain
\beq
(D-3)M-(D-2)\left(\sum_{q}J_{q}\Omega_q+\sum_{\hat q}\hat J_{\hat q}\hat\Omega_{\hat q}\right)-(D-3-q)\Phi_H Q_p=\mathcal{T}_{\text{tot}}~~,
\eeq
where the total tension $\mathcal{T}_{\text{tot}}$ is defined as
\beq
\mathcal{T}_{\text{tot}}=\int_{\mathcal{B}_p}dV_{(p)}\left((D-2)\mathcal{P}\gamma^{ab}\xi_b-\mathcal{P}^{a}_\xi+(q-1)\Phi_HJ^{a}_{q}\right)n_a~~.
\eeq
Here we have replaced the total charge by an integral over a conserved current. For the case $q=p$ one should replace the integral over $J^{a}_{q}n_a$ with the appropriate integral involving $J^{a_1...a_{p+1}}_q$. 
In the case in which the configurations describe asymptotically flat black holes we must have that $\mathcal{T}_{\text{tot}}$ vanishes.

\section{Constraints on spin transport coefficients}
\label{sec:constraints}

In this section we study the constraints on the spin transport coefficients appearing in the Lagrangian \eqref{lagrangian} using the framework presented in the previous sections. We can regard the Lagrangian \eqref{lagrangian} as a derivative expansion, counting the number of derivatives applied to $g_{\mu\nu}$ and $\gamma_{ab}$ (writing ${\bf k}^a$ in terms of $u^a$ and ${\bf k}$).
Write this derivative expansion as $\CL = \CL_0 + \CL_1 + \CL_2 + \CL_3 + \cdots$ where $\CL_m$ consists of the terms with $m$ derivatives. Hence by \eqref{thermodyn} we can make the following derivative expansion of the transverse angular momenta
\begin{equation}
\label{hatJ1}
\hat{J}_{\hat{q}} =   \int_{\CB_p} dV_{(p)} R_0 \left( \frac{\partial \CL_0}{\partial \hat{\Omega}_{\hat{q}}} + \frac{\partial \CL_1}{\partial \hat{\Omega}_{\hat{q}}} + \frac{\partial \CL_2}{\partial \hat{\Omega}_{\hat{q}}} + \cdots \right)~~.
\end{equation}
This identity employs the thermodynamic relations to compute $\hat{J}_{\hat{q}}$. One can equivalently compute $\hat{J}_{\hat{q}}$ using \eqref{ccharge} with the conserved current \eqref{trans_cur} for $\hat{\chi}_{\hat{q}}$.
We compute
\begin{equation}
\label{hatJ2}
\hat{J}_{\hat{q}} =  \int_{\CB_p} dV_{(p)} n_a \CS^{a \mu\nu} \nabla_\mu \hat{\chi}_{\hat{q},\nu} =  \int_{\CB_p} dV_{(p)} n_a \CS^{a ij} \epsilon^{(\hat{q})}_{ij} =  \int_{\CB_p} dV_{(p)} n_a \frac{\partial \CL}{\partial \omega_a^{(\hat{q})}} ~~.
\end{equation}
Here we have defined the spin current $\CS^a_{(\hat{q})}$ for the transverse spin plane associated with $\hat{\chi}_{\hat{q}}$ as
\begin{equation}
\CS^{aij} = \sum_{\hat{q}} \CS^a_{(\hat{q})} \epsilon^{ij}_{(\hat{q})} \spa \CS^a_{(\hat{q})} = \frac{1}{2} \frac{\partial \CL}{\partial \omega_a^{(\hat{q})}}~~,
\end{equation}
and we also translated the relation \eqref{Tab} of the spin current obtained as the derivative of the Lagrangian with respect to the extrinsic twist potential into the corresponding relation for the individual transverse spin planes. Moreover, in \eqref{hatJ1} we used the identity \eqref{imp_id} derived in Appendix \ref{trans_spin}.

Expanding now \eqref{hatJ2} in the number of derivatives and comparing this to \eqref{hatJ1} we can infer that we have the following relation%
\footnote{We assume here that \eqref{trans_cur} is true to all orders. When only including pole-dipole corrections we have shown that this is true up to and including $m=2$, but we have not included higher poles that could enter at $m=2$.}
\begin{equation}
\label{spin_constraints}
R_0  \frac{\partial \CL_m}{\partial \hat{\Omega}_{\hat{q}}} =  n_a \frac{\partial \CL_{m+1}}{\partial \omega_a^{(\hat{q})}} ~~,~~ m=0,1,2,... ~~.
\end{equation}
Thus, we have found a relation between the $m$-derivative terms and the $(m+1)$-derivative terms in the Lagrangian. This is a rather powerful set of constraints. For one thing, if we use them iteratively, one can conclude that the terms in the Lagrangian containing the extrinsic twist potential and with $m$ derivatives must involve a product of $m$ extrinsic twist potentials. We can write this as
\begin{equation}
\CL^{\rm (spin)}_m = \frac{1}{m!} \sum_{\hat{q},\hat{q}'} L^{a_1 \cdots a_m}_{\hat{q}_1,...,\hat{q}_m} \omega_{a_1}^{(\hat{q}_1)} \cdots \omega_{a_m}^{(\hat{q}_m)} ~~.
\end{equation}
In order for these terms to have precisely $m$ derivatives the tensors $L^{a_1 \cdots a_m}_{\hat{q}_1,...,\hat{q}_m}$ cannot contain any derivatives. The tensors $L^{a_1 \cdots a_m}_{\hat{q}_1,...,\hat{q}_m}$ are symmetric when exchanging $(a_i,\hat{q}_i) \leftrightarrow (a_j,\hat{q}_j)$. Applying \eqref{spin_constraints} repeatedly we find
\begin{equation}
\label{constr1}
\frac{1}{R_0^{m}} n_{a_1} \cdots n_{a_m} L^{a_1 \cdots a_m}_{\hat{q}_1,...,\hat{q}_m} = \frac{\partial^m \lambda_0}{\partial \hat{\Omega}_{\hat{q}_1} \cdots \partial \hat{\Omega}_{\hat{q}_m}} ~~,
\end{equation}
where we used $\CL_0=\lambda_0$ from \eqref{lagrangian}. This is obviously a very powerful set of constraints on the tensors $L^{a_1 \cdots a_m}_{\hat{q}_1,...,\hat{q}_m}$. However, it does not completely determine them. 

One can furthermore consider the equivalence between the conserved currents \eqref{long_cur} and the thermodynamics \eqref{thermodyn} for $M$ and $J_q$. This gives the following constraints for the spin terms in the Lagrangian
\begin{equation}
\xi^a V_a = 0 \spa \chi^a_q V_a = 0 \spa V_a \equiv 
2  \gamma_{ac} n_b \frac{\partial \CL_m^{\rm (spin)}}{\partial \gamma_{cb}} + R_0  \frac{\partial \CL_m^{\rm (spin)}}{{\bf k}^a} +  n_b \sum_{\hat{q}} \frac{\partial \CL^{\rm (spin)}_{m}}{\partial \omega_b^{(\hat{q})}} \omega_{a}^{(\hat{q})}  ~~.
\end{equation}
Using \eqref{spin_constraints} we find
\begin{equation}
\label{constr2}
V_a = 2  \gamma_{ac} n_b \frac{\partial \CL_m^{\rm (spin)}}{\partial \gamma_{cb}} + R_0  \frac{\partial \CL_m^{\rm (spin)}}{{\bf k}^a} + R_0 \frac{\partial \CL^{\rm (spin)}_{m-1}}{\partial \hat{\Omega}_{\hat{q}}} \omega_{a}^{(\hat{q})} ~~.
\end{equation}
Demanding $\xi^a V_a = \chi^a_q V_a = 0$ corresponds in practice to demanding $V_a=0$ since it is limited what tensors $L^{a_1, \cdots, a_m}_{\hat{q}_1,...,\hat{q}_m}$ one can write down without derivatives. 

We show now that by combining the constraint \eqref{constr1} and the constraint $V_a=0$ with \eqref{constr2} we can fix uniquely the spin terms $\CL_m^{(spin)}$ for $m=1,2$. For $m=1$ one could imagine writing $L^a_{\hat{q}} = \xi^a f_{\hat{q}}( {\bf k} )$ or $L^a_{\hat{q}} = {\bf k}^a f_{\hat{q}}( {\bf k} )$. However, the first proposal does not work since the first two terms in \eqref{constr2} are zero while the last term is not. Instead for the second proposal both \eqref{constr1} and \eqref{constr2} gives
\begin{equation}
\label{Lspin1}
\CL^{\rm (spin)}_1 = - {\bf k} u^a \sum_{\hat{q}} \frac{\partial\lambda_0}{\partial \hat{\Omega}_{\hat{q}}} \omega_a^{(\hat{q})} ~~.
\end{equation}
For $m=2$ one can imagine more possibilities. Consider the ansatz
\begin{equation}
\frac{1}{2} L^{ab}_{\hat{q},\hat{q}'} = \gamma^{ab} f_{\hat{q},\hat{q}'} ({\bf k}) + u^a u^b g_{\hat{q},\hat{q}'} ({\bf k}) ~~.
\end{equation}
The constraint \eqref{constr1} gives $- {\bf k}^2 f_{\hat{q},\hat{q}'} + R_0^2 g_{\hat{q},\hat{q}'} = \frac{1}{2} R_0^2 {\bf k}^2 \frac{\partial^2\lambda_0}{\partial \hat{\Omega}_{\hat{q}}\partial \hat{\Omega}_{\hat{q}'}}$.
Instead the constraint \eqref{constr2} gives 
\begin{equation}
0 = V_a = -2 n^b \sum_{\hat{q},\hat{q}'} f_{\hat{q},\hat{q}'} \omega_{a}^{(\hat{q})}  \omega_{b}^{(\hat{q}')} + \frac{2 R_0 u^b}{\bf k}  \sum_{\hat{q},\hat{q}'}  \left( g_{\hat{q},\hat{q}'} - \frac{1}{2} {\bf k}^2 \frac{\partial^2\lambda_0}{\partial \hat{\Omega}_{\hat{q}}\partial \hat{\Omega}_{\hat{q}'}} \right)  \omega_{a}^{(\hat{q})}  \omega_{b}^{(\hat{q}')} ~~.
\end{equation}
Hence the constraints \eqref{constr1} and \eqref{constr2} fix uniquely 
\begin{equation}
\label{Lspin2}
\CL^{\rm (spin)}_2 = \frac{{\bf k}^2}{2} u^a u^b \sum_{\hat{q},\hat{q}'} \frac{\partial^2\lambda_0}{\partial \hat{\Omega}_{\hat{q}}\partial \hat{\Omega}_{\hat{q}'}} \omega_a^{(\hat{q})}\omega_b^{(\hat{q}')}~~.
\end{equation}
It is interesting to consider the physical interpretation of the transverse spin terms \eqref{Lspin1} and \eqref{Lspin2}. Define
\begin{equation}
\label{angveleff}
\delta \hat{\Omega}_{\hat{q}} \equiv - {\bf k}^a \omega_a^{(\hat{q})}~~.
\end{equation}
Then we can generate the transverse spin terms \eqref{Lspin1} and \eqref{Lspin2} from the zeroth order term $\lambda_0$ in the Lagrangian \eqref{lagrangian} by substituting $\hat{\Omega}_{\hat{q}}$ with $\hat{\Omega}_{\hat{q}} + \delta \hat{\Omega}_{\hat{q}}$, $i.e.$ 
\begin{equation}
\CL = \lambda_0 (T,{\bf k} ,  \hat{\Omega}_{\hat{q}} + \delta \hat{\Omega}_{\hat{q}} ) +  \upsilon_1 \omega^{ab} \omega_{ab} + \upsilon_2 \CR + \upsilon_3 u^a u^b \CR_{ab} + \lambda_1 K^i K_i + \lambda_2 K^{abi} K_{abi} + \lambda_3 u^a u^b K_a{}^{ci}K_{bci}~~.
\end{equation}
Note that all the other terms are of second order in the number of derivatives hence they will not generate new terms to the order we are considering. The interpretation of this is that subjecting the brane to a background with non-zero extrinsic twist potential, corresponding to frame-dragging terms for the transverse spin planes, adds an effective angular velocity $\delta \hat{\Omega}_{\hat{q}}$ defined in \eqref{angveleff} to the leading order angular velocity $\hat{\Omega}_{\hat{q}}$  of the brane. Hence, even if we start with a fluid brane embedding with zero angular velocity in the transverse spin planes, to zeroth order, the frame-dragging effects of the background can provide an effective non-zero angular velocity in the transverse spin planes.

\section{Application to higher-dimensional black holes}\label{sec:examples}
In this section we apply the thermodynamic formulas of the previous sections to the perturbative construction of higher-dimensional black holes. The case of neutral black rings in asymptotically flat space has been considered in \cite{Emparan:2007wm, Armas:2013hsa, Armas:2014bia}. Here we take into account the second order corrections in \eqref{gibbsfreeenergy} in the case of neutral black tori, as well as black rings carrying Maxwell charge ($q=0$) in Einstein-Maxwell-Dilaton gravity. The results obtained for this charged configuration are compared to the full numerical solution in $D=7$, which we also obtain via a solution generating technique taking the neutral black ring numerically found in \cite{Dias:2014cia} as the seed solution. The agreement that will be verified is already expected due to results obtained for the neutral case \cite{Armas:2014bia}. However, this comparison is an important check of our effective theory of charged black branes and in the process, new families of charged black rings are numerically generated. 

\subsection{Black tori}
 Black tori were constructed in the perfect fluid limit in \cite{Emparan:2009vd} and are black hole objects with $\mathbb{T}^{p}\times s^{n+1}$ horizon geometry in pure Einstein gravity. In the case $p=1$ they correspond to the black rings studied in \cite{Emparan:2007wm, Armas:2013hsa, Armas:2014bia}. In order to embed this geometry into $D$-dimensional flat space we consider the Minkowski metric written in the form
 \beq \label{flatm}
 ds^2=-dt^2+\sum_{\hat{a}=1}^{p}\left(dr_{\hat{a}}^2+r_{\hat{a}}^2d\theta_{\hat{a}}^2\right)+\sum_{i=1}^{n-p+2}dx_{i}^2~~,
 \eeq
 with $\hat{a}=1...p$ and choose the embedding coordinates
 \beq\label{tc}
 t=\tau~~,~~\theta_{\hat{a}}=\phi_{\hat{a}}~~,~~r_{\hat{a}}=R_{\hat{a}}~~,~~x_i=0~~.
 \eeq
 The resulting induced metric is thus manifestly flat. We consider a geometry which is rotating with different angular velocities $\Omega_{\hat{a}}$ along each of the angular coordinates $\phi_{\hat{a}}$ so that the horizon Killing vector field is
 \beq
 \textbf{k}^{a}\partial_a=\partial_\tau+\sum_{{\hat{a}}=1}^{p}\Omega_{\hat{a}}\partial_{\phi_{{\hat{a}}}}~~,
 \eeq
 and has a constant norm $\textbf{k}$ along the $(p+1)$-dimensional world-volume.
 
 We now consider the free energy \eqref{gibbsfreeenergy} for this specific configuration. Since the world-volume is flat, the fluid flows have no vorticity and we are not considering transverse spin, the only relevant corrections are those parametrised by the coefficients $\lambda_1,\lambda_2,\lambda_3$. However, due to Gauss-Codazzi relations in a flat background, these configurations, as black rings, are only described by one single transport coefficient $\tilde\lambda_1=\lambda_1+\lambda_2+(1/n)\lambda_3$ of the form \cite{Armas:2013hsa, Armas:2013goa}\footnote{The measurement of the transport coefficients $\lambda_1,\lambda_2,\lambda_3$ by elastically perturbing black branes was made in \cite{Armas:2011uf, Camps:2012hw}.}
\beq
\tilde\lambda_{1}=-Pr_0^2\frac{(n+1)(3n+4)}{2n^2(n+2)}\xi(n)~~,~~\xi(n)=\frac{n\tan(\pi/n)}{\pi}\frac{\Gamma\left(\frac{n+1}{n}\right)^4}{\Gamma\left(\frac{n+2}{n}\right)^2}~~,~~n\ge3~~,
\eeq
where $r_0$ is the local horizon radius of the black brane and $P=\lambda_0$ the pressure given by \cite{Emparan:2009at}
\beq \label{pressure}
P=-\frac{\Omega_{(n+1)}}{16\pi G}r_0^n~~,~~r_0=\frac{n}{4\pi T}\textbf{k}~~,
\eeq
 with $G$ being Newton's constant and $\Omega_{(n+1)}$ the unit volume of a $(n+1)$-dimensional sphere. The free energy then becomes
 \beq \label{freetori}
 \mathcal{F}[R_{\hat{a}}]=-V_{(p)}\left(P+\tilde\lambda_1K^{i}K_{i}\right)~~,
 \eeq
 where the square of the mean extrinsic curvature and the volume of the $p+1$ brane are given by
 \beq
 K^{i}K_{i}=-\sum_{{\hat{a}}=1}^{p}\frac{1}{R_{\hat{a}}^2}~~,~~V_{(p)}=(2\pi)^{p}\prod_{{\hat{a}}=1}^{p}R_{{\hat{a}}}~~.
 \eeq
The free energy \eqref{freetori} is the higher-codimension version of the Helfrich-Canham free energy of biophysical membranes \cite{Armas:2014bia}. From \eqref{freetori} follows a set of $p$ coupled equations, each obtained by varying \eqref{freetori} for each of the radii $R_{{\hat{a}}}$. The solution to this set of equations takes the simple form
 \beq \label{soltori}
 \Omega_{{\hat{a}}}R_{\hat{a}}=\Omega_{(0){\hat{a}}}R_{\hat{a}}\left(1+\frac{(n+1)(3n+4)}{2n^2(n+2)}\xi(n)\tilde\varepsilon^2_{\hat{a}}\right)~~,
 \eeq
 where we have defined the parameters $\tilde\varepsilon^2_{\hat{a}}=r_0^2/R_{\hat{a}}^2$ and used the leading order result $\Omega_{(0){\hat{a}}}$ previously obtained in \cite{Emparan:2009vd}
 \beq
 \Omega_{(0){\hat{a}}}R_{\hat{a}}=\frac{1}{\sqrt{n+p}}~~.
 \eeq
 In the case $p=1$ the result \eqref{soltori} reduces to that obtained in \cite{Armas:2013hsa, Armas:2014bia}. We now evaluate the thermodynamic properties of these configurations using the formulas of Sec.~\ref{thermo}. These are given by
 \beq \label{mt1}
 M=\frac{V_{(p)}}{16\pi G}r_0^n R_{\hat{a}}\sqrt{n+p}\left(1-\frac{(n+1)(3n+4)}{2n^2(n+2)}\xi(n)\sum_{{\hat{a}}=1}^{p}\tilde\varepsilon^2_{\hat{a}}\right)~~,
 \eeq
 \beq
 J_{\hat{a}}=\frac{V_{(p)}}{16\pi G}r_0^n R_{\hat{a}}\sqrt{n+p}\left(1-\frac{(n+1)(3n+4)}{2n^2(n+2)}\xi(n)\!\!\!\!\!\!\sum_{{\hat{b}}=1/\{{\hat{b}}={\hat{a}}\}}^{p}\!\!\!\!\!\!\ \tilde\varepsilon^2_{\hat{b}}\right)~~,
 \eeq
  \beq\label{mt3}
S=\frac{V_{(p)}}{4 G}r_0^{n+1} \sqrt{\frac{n+p}{n}}\left(1-\frac{(n+1)^3(3n+4)}{2n^3(n+2)(n+p)}\xi(n)\sum_{{\hat{a}}=1}^{p}\tilde\varepsilon^2_{\hat{a}}\right)~~.
 \eeq
It is straightforward to check that when $p=1$ these thermodynamic properties reduce to those obtained in \cite{Armas:2014bia} and that in the leading order case, where $\tilde\varepsilon_{\hat{a}}=0$, we obtain the results presented in \cite{Emparan:2009vd}.

\subsection{Black rings with Maxwell charge}
We now consider the case of black rings carrying Maxwell charge in Einstein-Maxwell-Dilaton theory with Kaluza-Klein coupling parameter. In the perfect fluid limit, these solutions have been constructed in \cite{Caldarelli:2010xz}. We compute the corrections due to the transport coefficients $\lambda_1,\lambda_2,\lambda_3$ and compare the results in $D=7$ with solutions generated numerically using the numerical neutral black ring solution \cite{Dias:2014cia} as the starting point. In order to do so, we uplift the neutral numerical solution to $D=8$ along the direction $z$, boost it along the $(t,z)$ plane and reduce it over the $z$-direction. The resulting solution is a charged black ring and the details of this procedure are given in App.~\ref{nbr}.

As in the case of neutral black tori of the previous section we write the flat background metric as in \eqref{flatm} and choose the coordinates \eqref{tc} with $p=1$. The resulting induced metric is again manifestly flat. As in the case of the black tori, these geometries are only parametrised by a single transport coefficient  $\tilde\lambda_1=\lambda_1+\lambda_2+(1/n)\lambda_3(1-\Phi_H^2/\textbf{k}^2)$ given by \footnote{The transport coefficients $\lambda_1,\lambda_2,\lambda_3$ for charged black branes obtained via Kaluza-Klein reduction were obtained in \cite{Armas:2012ac, Armas:2013aka}.} 
\beq
\tilde\lambda_1=- P\frac{(3n+4)}{2n^2(n+2)}\left(n+1+\frac{\Phi_H^2}{\textbf{k}^2-\Phi_H^2}\right)\xi(n) r_0^2~~,~~r_0=\frac{n}{4\pi T}\textbf{k}\sqrt{1-\frac{\Phi_H^2}{\textbf{k}^2}}~~,
\eeq
where $P$ was defined in \eqref{pressure}. The free energy therefore takes the same form as in \eqref{freetori} for $p=1$. Varying it and solving the respective equation of motion we find a balancing condition for $\Omega$ of the form
\beq
\Omega=\Omega_{(0)}\left(1+\frac{(3n+4)(1+n(1-\Phi_H^2))}{2n^2(n+2)(1-\Phi_H^2)}\xi(n)\tilde\varepsilon^2\right)~~,~~\Omega_{(0)}=\frac{1}{R}\sqrt{\frac{1-\Phi_H^2}{(n+1)}}~~,
\eeq
where $R=R_1$ and $\Omega=\Omega_{1}$. The leading order solution $\Omega_{(0)}$ was previously obtained in \cite{Caldarelli:2010xz}. In the case where the charge vanishes $\Phi_H\to0$ we obtain the result given in \eqref{soltori} when $p=1$.

We can use the thermodynamic formulas of Sec.~\ref{thermo} and Sec.~\ref{charge} to obtain the thermodynamic quantities of these solutions. These take the form
\beq
M=\frac{\Omega_{(n+1)}r_0^n}{8G} R\frac{(n+2-\Phi_H^2)}{1-\Phi_H^2}\left(1-\frac{(n+1)(3n+4)(n(n+2)(1-\Phi_H^2)+(n+1)\Phi_H^4)}{2n^3(n+2)(1-\Phi_H^2)}\xi(n)\tilde\varepsilon^2\right)~~,
\eeq
\beq
J=\frac{\Omega_{(n+1)}r_0^n}{8G}\frac{\sqrt{n+1}}{\sqrt{1-\Phi_H^2}}R^2\left(1+\frac{(3n+4)\Phi^2_H}{n^3(n+2)(1-\Phi_H^2)}\xi(n)\tilde\varepsilon^2\right)~~,
\eeq
\beq
S=\frac{\Omega_{(n+1)}r_0^{n+1}}{2G}\pi R \sqrt{\frac{n+1}{n(1-\Phi_H^2)}}\left(1-\frac{(3n+4)(n(n+1)^2-(-2+n(-1+n+n^2))\Phi_H^2)}{2n^4(n+2)(1-\Phi_H^2)}\xi(n)\tilde\varepsilon^2\right)~~,
\eeq
\beq
Q_{(1)}=\frac{\Omega_{(n+1)}r_0^n}{8G} R(n+1)\frac{\Phi_H}{1-\Phi_H^2}\left(1-\frac{(n^2-1)(3n+4)(2-\Phi^2_H+n(1-\Phi_H^2))}{n^3(n+2)(1-\Phi_H^2)}\xi(n)\tilde\varepsilon^2\right)~~.
\eeq
These quantities in the limit $\Phi_H\to0$ reduce to those obtained in \eqref{mt1}-\eqref{mt3} for $p=1$ and when $\tilde\varepsilon=0$ to those obtained in \cite{Caldarelli:2010xz}.

\subsubsection*{Phase diagram}
Given the asymptotic charges of the solution we can construct the phase diagram of the black rings and compare it to the full numerical solution as in the case of neutral black rings \cite{Armas:2014bia}. In order to do so we introduce the reduced dimensionless thermodynamic charges as in \cite{Emparan:2007wm},
\beq \label{reducedthermo}
j^{n+1}=c_j \frac{J^{n+1}}{GM^{n+2}}~~,~~a_{\text{H}}^{n+1}=4^{n+1}c_a\frac{S^{n+1}}{(GM)^{n+2}}~~,~~\omega_{\text{H}}=c_\omega \Omega(GM)^{\frac{1}{n+1}}~~,~~t_{\text{H}}=c_t T (GM)^{\frac{1}{n+1}}~~,
\eeq
where, 
\beq
c_j=\frac{(16\pi)^{n+1}}{2^{n+4}n^{\frac{n+1}{2}}}c_a=\frac{\Omega_{(n+1)}}{2^{n+5}}\frac{(n+2)^{n+2}}{(n+1)^{\frac{n+1}{2}}}~~,~~c_{\omega}=\frac{\sqrt{n}}{4\pi}8^{\frac{1}{n+1}}c_t=\sqrt{n+1}\left(\frac{n+2}{16}\Omega_{(n+1)}\right)^{-\frac{1}{n+1}}.~~
\eeq
In order to derive these quantities and their relations analytically it is useful to use the freedom of shifting $R$ by a small amount (see App.~\ref{cc}) in order to find a choice of surface, parametrised by $R$, which renders $j$ without any $\tilde\varepsilon^2$ corrections. This is achieved by performing the transformation $R\to\nu\xi(n)r_0 \tilde\varepsilon$ such that
\beq
\nu=\frac{(3n+4)(n(n+2)^2-(n+2)(-1+n(n+2))\Phi_H^2+(-3+n+n^2)\Phi_H^4)}{2n^3(n+2)(2+n-\Phi_H^2)(-1+\Phi_H^2)}~~.
\eeq
Given this redefinition of $R$ we find the following form for the reduced angular momentum $j$
\beq \label{je}
j=\tilde\varepsilon^{-\frac{n}{n+1}}\frac{2^{-\frac{n+2}{n+1}}}{\sqrt{1-\Phi_H^2}}\left(\frac{n+2-\Phi_H^2}{(n+2)(1-\Phi_H^2)}\right)^{-\frac{n+2}{n+1}}~~.
\eeq
The parameter $\tilde\varepsilon$ counts the order of the perturbative expansion and one must require $\tilde\varepsilon\ll1$. From \eqref{je} we see that this implies $j\gg1$. However, we will see shortly that our results can be successfully extrapolated to the regime $j\sim\mathcal{O}(1)$, as expected due to the results for the neutral case \cite{Armas:2014bia}. The result \eqref{je} when introduced in the remaining reduced quantities leads to the following form of the phase diagram expressed in terms of $j$ and $\Phi_H$,
\beq \label{p1}
a_H(j,\Phi_H)=\frac{2^{\frac{n-2}{n(n+1)}}}{j^{\frac{1}{n}}}\frac{(1-\Phi_H^2)^{\frac{n+3}{2n}}}{\left(n+2-\Phi_H^2\right)^{\frac{n+2}{n}}}\left(n+2\right)^{\frac{n+2}{n}}\left(1+\frac{(n+1)(3n+4)(1-\Phi_H^2)^{\frac{3}{n}}}{2^{\frac{3n+4}{n}}n^4(n+2-\Phi_H^2)^{\frac{3n+4}{n}}}\frac{f_a(n,\Phi_H)}{j^{2\frac{(n+1)}{n}}}\right)~~,
\eeq
\beq
t_{\text{H}}(j,\Phi_H)=\frac{2^{\frac{2-n}{n(n+1)}}}{(n+2)^{\frac{2}{n}}} n j^{\frac{1}{n}} \frac{(n+2-\Phi_H^2)^{\frac{2}{n}}}{(1-\Phi_H^2)^{\frac{3}{2n}}}\left(1+\frac{(n+1)(3n+4)(1-\Phi_H^2)^{\frac{3}{n}}}{2^{\frac{3n+4}{n}}n^4(n+2-\Phi_H^2)^{\frac{3n+4}{n}}}\frac{f_{t}(n,\Phi_H)}{j^{\frac{2(n+1)}{n}}}\right)~~,
\eeq
\beq
\omega_{\text{H}}(j,\Phi_H)=\frac{(n+2)}{2 j}\frac{1-\Phi_H^2}{(n+2-\Phi_H^2)}\left(1+\frac{(n+1)(3n+4)(1-\Phi_H^2)^{\frac{6-n}{2n}}}{2^{\frac{2(n+2)}{n}}n^3(n+2-\Phi_H^2)^{\frac{3n+4}{n}}}\frac{f_{\omega}(n,\Phi_H)}{j^{\frac{2(n+1)}{n}}}\right)~~,
\eeq
where,
\beq\label{p4}
\begin{split}
f_a(n,\Phi_H)&=(n+2)^{\frac{n+4}{n}}\left(n(n+2)-(2+n(n+4))\Phi_H^2+(n-1)\Phi_H^4\right)\xi(n)~~,\\
f_{t}(n,\Phi_H)&=(n+2)^{\frac{n+4}{n}}\left(-3n(n+2)+(-2+3n(n+2))\Phi_H^2-3(n-1)\Phi_H^4\right)\xi(n)~~,\\
f_{\omega}(n,\Phi_H)&=(n+2)^{\frac{n+4}{n}}\left(n(n+2)+(1-n(n+2))\Phi_H^23(n-1)\Phi_H^4\right)\xi(n)~~.
\end{split}
\eeq
In the case $\Phi_H\to0$ these relations result in those presented in \cite{Armas:2014bia}. We now use these expressions and compare it to the numerical charged solutions obtained in App.~\ref{nbr} and the charged Myers-Perry solutions obtained in \cite{Kunz:2006jd} from Kaluza-Klein reduction. The reduced thermodynamic quantities \eqref{reducedthermo} for these solutions can be obtained from the asymptotic charges presented in \cite{Kunz:2006jd}. 

Below we plot the reduced area as a function of $j$ and $\Phi_H$ for the values $\Phi_H=0,~\Phi_H=\tanh0.5,~\Phi_H=\tanh0.8$. The solid thick lines correspond to the charged Myers-Perry solutions, while the solid thin lines to charged numerical solutions obtained using App.~\ref{nbr}. The thin dashed lines represent the phase diagram obtained via Eqs.~\eqref{p1}-\eqref{p4}. The thin dashed cyan line is the phase diagram obtained for $\Phi_H=\tanh0.5$ in the absence of corrections, that is, when we ignore the terms proportional to $\xi(n)$.

\begin{figure}[H]
\centering
  \includegraphics[width=0.4\linewidth]{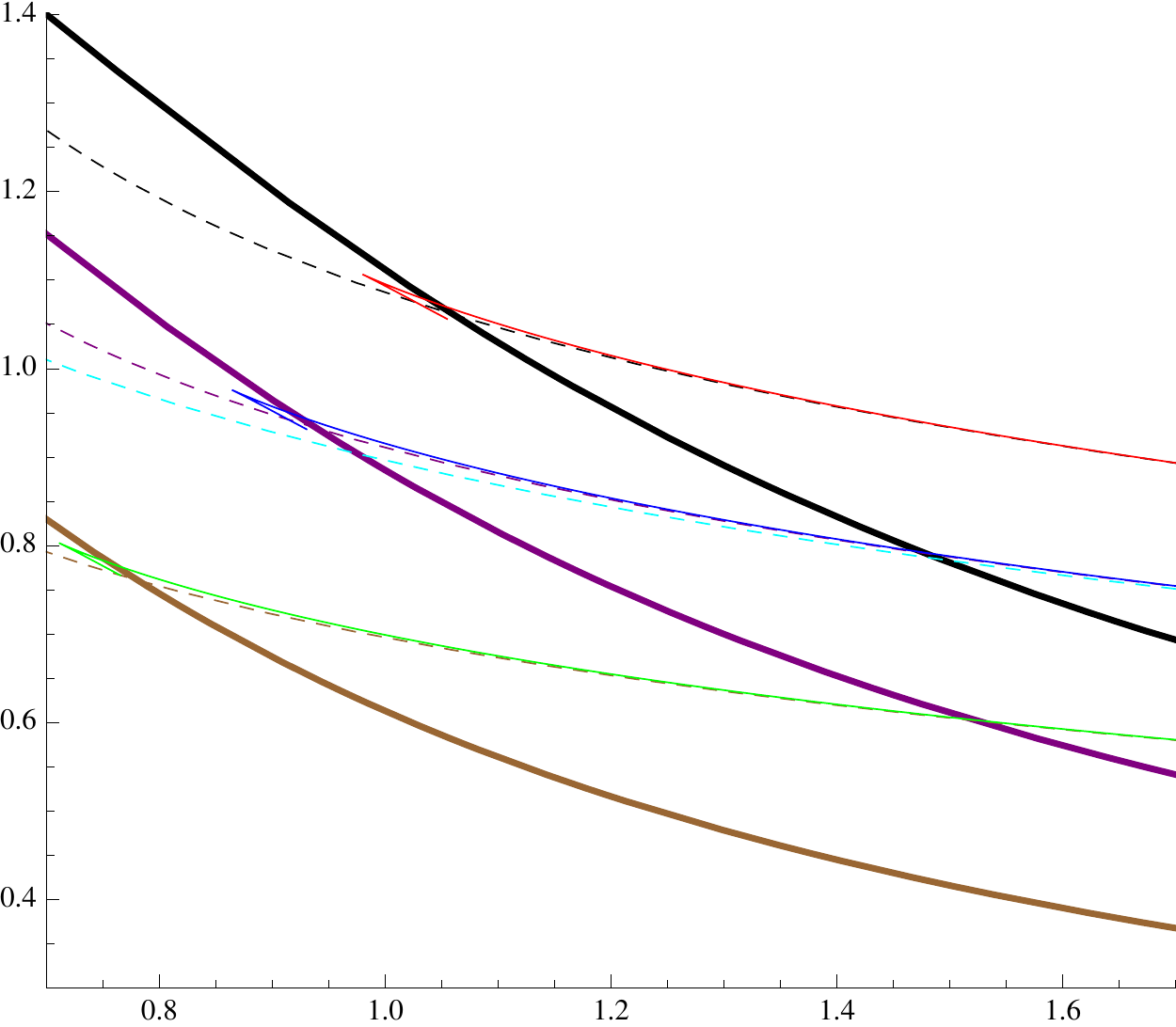}
  \begin{picture}(0,0)(0,0)
\put(-220,160){ $ a_{H}  $}
\put(-20,-5){ $ j$}
\end{picture}	
\caption{The reduced area $a_H$ in $D=7$ as a function of $j,\Phi_H$. The thick solid lines correspond to the charged Myers-Perry black holes for $\Phi_H=0$ (black line), $\Phi_H=\tanh0.5$ (purple line), $\Phi_H=\tanh0.8$ (brown line). The dashed lines of the same colour correspond to the curves obtained using  Eqs.~\eqref{p1}-\eqref{p4} for the same values of $\Phi_H$. The cyan curve is the infinitely thin approximation for $\Phi_H=\tanh0.5$ obtained in \cite{Caldarelli:2010xz}. The solid thin lines correspond to the numerically constructed black rings using App.~\ref{nbr}, in particular, the solid thin red line represents $\Phi_H=0$, the blue line $\Phi_H=\tanh0.5$ and the green line $\Phi_H=\tanh0.8$. }
\end{figure}
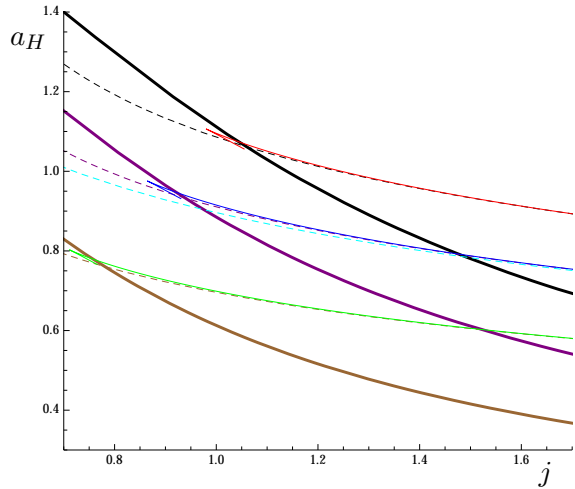
We can see that the agreement with the numerical solutions is striking for all values of $\Phi_H$, as expected due to the results for the neutral case \cite{Armas:2014bia}, and that it gives a much better approximation than in the case where corrections are ignored. This provides a check that the effective theory formulated here for charged black branes is correct. As in the neutral black ring case, we see that charged black rings tend to increase their reduced area (entropy) for a given value of $j$ and $\Phi_H$ compared to the case where terms proportional to $\xi(n)$ are ignored. Below we also exhibit the behaviour of the reduced temperature and angular velocity.
\begin{figure}[H]
\centering
\begin{subfigure}{.5\textwidth}
  \centering
  \includegraphics[width=0.8\linewidth]{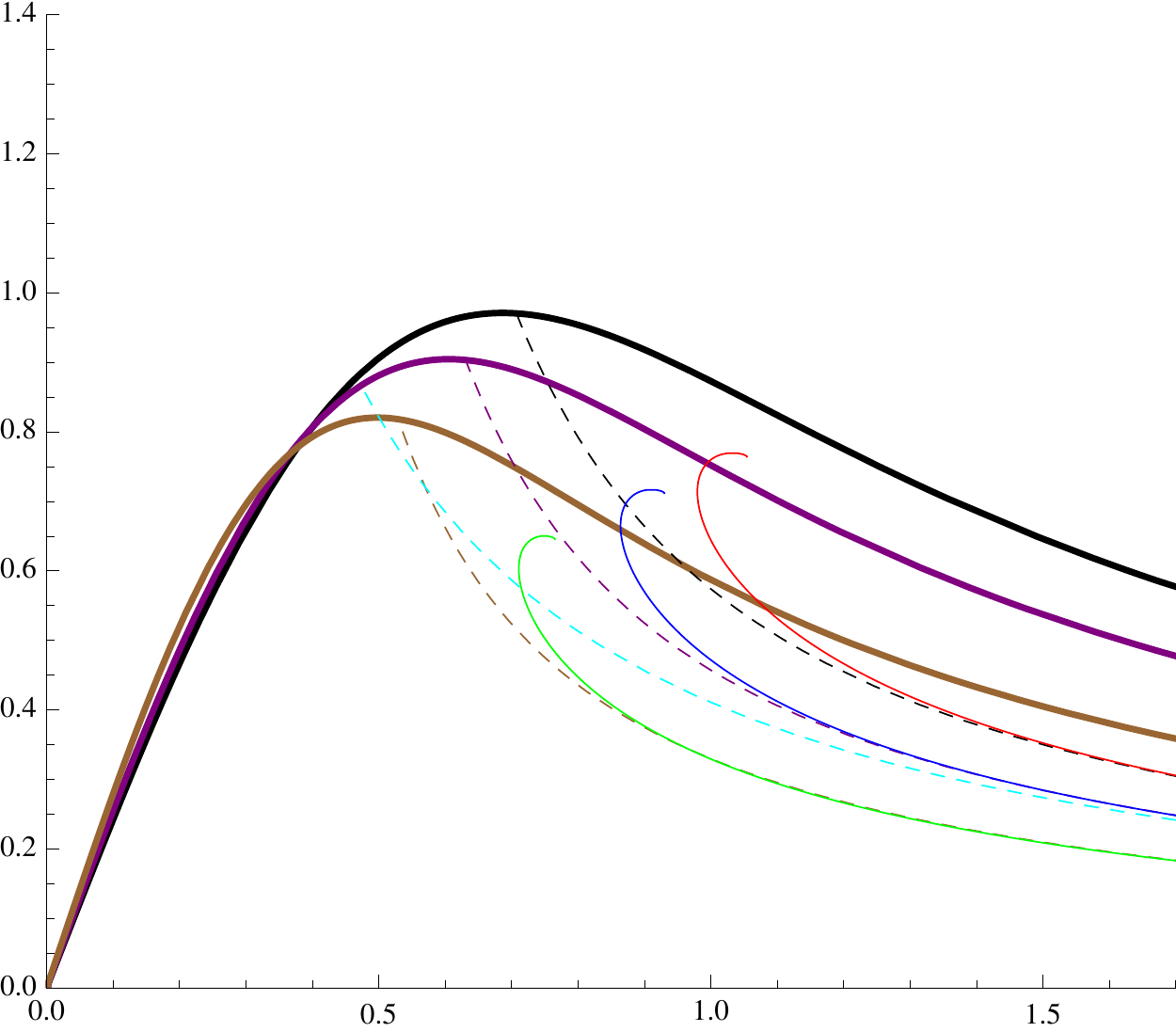}
  \begin{picture}(0,0)(0,0)
\put(-230,150){ $ \omega_H  $}
\put(-20,-5){ $ j$}
\end{picture}	
\end{subfigure}%
\begin{subfigure}{.5\textwidth}
  \centering
  \includegraphics[width=0.8\linewidth]{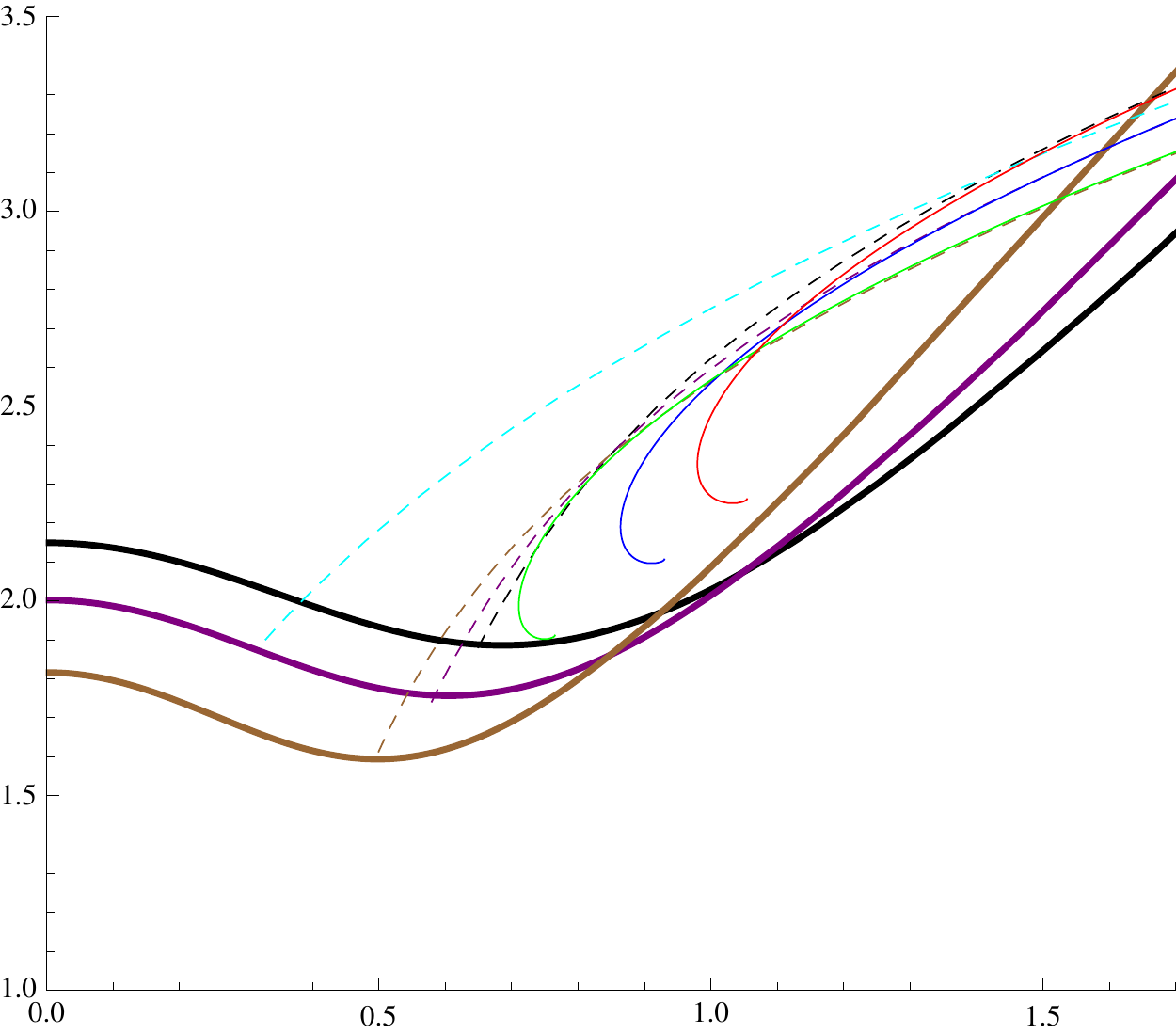}  
  \begin{picture}(0,0)(0,0)
  \put(-230,150){ $ t_{H}  $}
  \put(-20,-5){ $ j $}
  \end{picture}	
\end{subfigure}
\caption{On the left we have the behaviour of the reduced angular velocity and on the right of the reduced temperature. The colour coding is the same as in the previous figure.}
\end{figure}
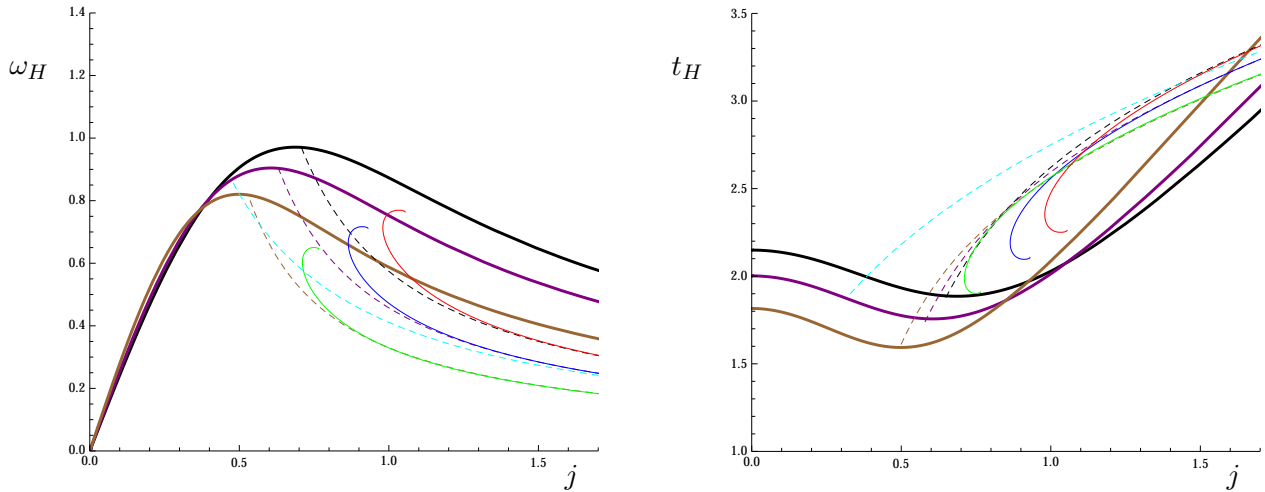
As mentioned below \eqref{je}, our analysis is valid for $\tilde\varepsilon\ll1$, for which the corrections are small, meaning that we must have $j\gg1$. However, we see that the agreement with the numerical solutions is excellent even when $j\sim\mathcal{O}(1)$, which was already guaranteed due to the fact that the same agreement is observed in the neutral case \cite{Armas:2014bia}. These charged black rings were conjectured to suffer from a Gregory-Laflamme instability in \cite{Caldarelli:2010xz}. Since we can see that introducing a small correction makes the approximation better, we are lead to confirm that these black rings are indeed unstable.

\section{Conclusions} \label{conclusions}
In this paper we have studied the thermodynamic properties of stationary fluid branes, with dynamics described by an effective action characterised by transport coefficients, and applied the results to higher-dimensional black holes, where the transport coefficients are measured via the blackfold approach. In particular, starting with the free energy, obtained via Wick rotation of the effective action, we have deduced different formulas for the thermodynamic quantities characterising fluid branes. A set of these formulas, presented in Eq.~\eqref{thermodyn}, is given in terms of the Killing vector field $\textbf{k}^{a}$ characterising the stationary fluid brane configuration. When the description is applied to black holes, $\textbf{k}^{a}$ is interpreted as the horizon Killing vector field and hence formulas \eqref{thermodyn} give a better physical picture in terms of the symmetries of the brane and background geometry. 

We have also analysed the constraints that followed by requiring these thermodynamic formulas, obtained under the principle that \eqref{gibbsfreeenergy} describes the generating function of the fluid brane, to match the global charges obtained by integrating conserved surface currents in the corresponding world-volume theory. This requirement constrains the transport coefficients $\upsilon_i,\lambda_i,\varpi_i$ to be functions of the local thermodynamic potentials of the fluid brane such as $\mathcal{T},\Omega'_{\hat q},\Phi_H$, which is expected from a hydrodynamics perspective. However, in the case of the transport coefficients associated with transverse angular momentum $\varpi_i$ this requirement imposes powerful constraints in the transport coefficients as seen in Sec.~\ref{sec:constraints}. In particular, if one neglects quadrupole and higher moments, one is able to find all the corrections to the effective action \eqref{fullaction} to an arbitrary order given only the leading order action in terms of $\lambda_0$.

The results of our analysis were generalised to the case of fluid branes carrying Maxwell or $q=p$-brane charge in the absence of external background fields, though it would be interesting to generalise the analysis to the case in which these fields are present. In the process, we gave generic expressions for the entropy and charge currents characterising stationary fluid branes as well as Smarr-type relations that such configurations need to satisfy. These results have allowed for a comparison between equilibrium partition functions and entropy current in \cite{Armas:2013goa} as well as to deduce the corrected phase diagram of higher dimensional neutral black rings \cite{Emparan:2007wm, Armas:2014bia}. Our analysis is general and does not depend on the particular choice of transport coefficients appearing in \eqref{lagrangian} but is instead applicable to all physical systems describable in terms of the effective action \eqref{fullaction}, which includes systems usually found in the context of biophysical membranes.

The understanding of the thermodynamics of these actions has allowed us, in the context of higher-dimensional black holes, to compute corrections to black tori in Einstein gravity and charged black rings in Einstein-Maxwell-Dilaton gravity within the blackfold approach in Sec.~\ref{sec:examples}. In the case of charged black rings we derived the full corrected phase diagram and compared it to charged Myers-Perry black holes in this theory as well as to full numerical black ring solutions in $D=7$, which we constructed via a solution generating technique and taking the neutral numerical black ring solution \cite{Dias:2014cia} as the starting point. We found that our effective description provides an excellent approximation to the full solution and that the agreement extends way beyond its regime of validity. This is a strong indication that these black holes suffer from a Gregory-Laflamme instability.

The results obtained in this paper can also be potentially applied in the context of thermal probes and the AdS/CFT correspondence \cite{Grignani:2010xm, Grignani:2012iw, Armas:2012bk, Niarchos:2012pn, Niarchos:2013ia, Armas:2013ota, Armas:2014nea}. In some of these cases, this would require a generalisation of our analysis to include possible couplings to background fields and, generically, since these geometries are not characterised by flat world-volumes, it would be necessary to push the analysis of \cite{Emparan:2007wm, Camps:2012hw} to next order in a derivative expansion and to measure the transport coefficients in \eqref{lagrangian} denoted by $\upsilon_{i}$ for $i=1,2,3$.

Finally, our analysis opens up different avenues of research. In particular, it has given us the necessary tools to extend the blackfold approach to the case of multi-spinning black branes such as the Kerr brane. This allows us to construct several new black hole solutions with multiple disconnected horizons such as doubly-spinning black rings in a doubly-spinning Myers-Perry black hole background and to analyse their stability. This will be the subject of a later publication \cite{Armas:new}. This analysis also allows for a deeper study between spinning black holes and current anomalies in charged fluid dynamics as well as entanglement entropy in theories with gravitational Chern-Simons terms \cite{Castro:2014tta}. We hope to address some of these questions in the future.

\section*{Acknowledgements}
We thank Roberto Emparan and Amos Yarom for useful discussions. We also thank \'{O}scar J.C. Dias, Jorge E. Santos and Benson Way for providing the numerical data that allowed us to construct numerical charged black rings. JA thanks Amos Yarom for hospitality at \textbf{Technion}, where part of this work was developed. JA also thanks the organisers of the workshop \textbf{New Frontiers in Dynamical Gravity} at the University of Cambridge (2014) for allowing the presentation of some of these results. JA has been supported by a short visit grant from the Holograv Network under the programme \textbf{Holographic Methods for Strongly Coupled Systems} and by the Swiss National Science Foundation and the �Innovations- und Kooperationsprojekt C-13� of the Schweizerische Universit\"{a}tskonferenz SUK/CUS. TH acknowledges support from the ERC-advance grant \textbf{Exploring the Quantum Universe} as well as from the Marie-Curie-CIG grant \textbf{Quantum Mechanical Nature of Black Holes} both from the European Union.

\appendix

\section{World-volume currents from space-time currents } \label{cc}
In this appendix we show how to obtain the conserved world-volume currents \eqref{conc} identified in \cite{Armas:2013hsa} from conserved space-time currents. The pole-dipole object is characterised by a symmetric space-time stress-energy tensor of the form\footnote{This is a generalisation, introduced in \cite{Armas:2013goa}, of the stress-energy tensor defined in \cite{Vasilic:2007wp}.}
\beq \label{pdstress}
\hat T^{\mu\nu}(x^{\alpha})\!=\!\int_{\mathcal{W}_{p+1}}\!\!\!\!\!\!\!d^{p+1}\sigma\sqrt{-\gamma}\left(B^{\mu\nu}(X^{\alpha}(\sigma^{a}))\frac{\delta^{D}(x^{\alpha}-X^{\alpha}(\sigma^a))}{\sqrt{-g}}-\nabla_{\rho}\left(B^{\mu\nu\rho}(X^{\alpha}(\sigma^{a}))\frac{\delta^{D}(x^{\alpha}-X^{\alpha}(\sigma^a))}{\sqrt{-g}}\right)\right)~.
\eeq
Therefore for any Killing vector field in the background space-time $k_{\mu}(x^{\alpha})$ one has a conserved space-time current of the form
\beq
P_{k}^{\nu}=\hat{T}^{\mu\nu}k_{\mu}~~,
\eeq
which satisfies the conservation equation
\beq \label{tosolve}
\nabla_{\nu}P_{k}^{\nu}=0~~,
\eeq
due to the symmetry properties of $\hat T^{\mu\nu}$ and the Killing equation. To obtain the conserved world-volume currents and corresponding charges one solves  Eq.~\eqref{tosolve} by introducing an arbitrary space-time function $f(x^{\alpha})$ of compact support, i.e., we solve
\beq
\int d^{D}x\sqrt{-g}\nabla_{\nu}P_{k}^{\nu}\thinspace f=0~~.
\eeq
Solving this equation is reminiscent to solving the conservation equation for a space-time particle current as in \cite{Armas:2012ac, Armas:2013aka}, but it differs from it since this particular current involves the stress-energy tensor, which solves its own conservation equation, and a space-time Killing vector, which solves the Killing equation. 

A series of partial integrations using \eqref{pdstress} leads to an integral of the form
\beq \label{integral}
\int_{\mathcal{W}_{p+1}}\!\!\!\!\!\!\!d^{p+1}\sigma\sqrt{-\gamma}\left(B^{\nu\mu}k_{\mu}\nabla_{\nu}f+B^{\nu\mu\rho}\nabla_{\rho}k_{\mu}\nabla_{\nu}f+B^{\nu\mu\rho}k_{\mu}\nabla_{\rho}\nabla_{\nu}f\right)=0~~.
\eeq
This integral is non-trivially invariant under the field redefinitions $X^{\mu}(\sigma^{a})\to X^{\mu}(\sigma)+\tilde\varepsilon^{\mu}(\sigma^{a})$, where $\tilde\varepsilon^{\mu}(\sigma^{a})$ is an infinitesimal displacement vector of perturbative order $\mathcal{O}(\tilde\varepsilon)$. To precisely see the invariance of this integral, note that we have the following transformations rules under the field redefinitions
\beq \label{extra2}
\delta_2\sqrt{-\gamma}=\sqrt{-\gamma}u^{a}_{\rho}\nabla_{a}\tilde\varepsilon^{\rho}~~,\delta_2B^{\mu\nu}=-B^{\mu\nu}u^{a}_{\rho}\nabla_{a}\tilde\varepsilon^{\rho}-2B^{\lambda(\mu}\Gamma^{\nu)}_{\lambda\rho}\tilde\varepsilon^{\rho} - \frac{\partial B^{\mu\nu}}{\partial X^{\rho}}\tilde\epsilon^{\rho}~~,
\eeq
\beq\label{extra21}
\delta_2B^{\mu\nu\rho}=-B^{\mu\nu}\tilde\varepsilon^{\rho}~~,~~\delta_2 f=\tilde\varepsilon^{\rho}\nabla_{\rho}f~~,~~\delta_2 k_{\mu}=\tilde\varepsilon^{\rho}\nabla_{\rho}k_{\mu}+k_{\rho}\nabla_{\mu}\tilde\varepsilon^{\rho}~~.
\eeq
In order to solve the integral \eqref{integral} we decompose $f$ and its derivatives in independent components such that \cite{Armas:2012ac, Armas:2013aka}
\beq
\nabla_{\mu}f=f_{\mu}^{\perp}+u^{a}_{\mu}\nabla_{a}f~~,~~\nabla_{\nu}\nabla_{\mu}f=f_{\mu\nu}^{\perp}+2f^{\perp}_{(\mu a}u^{a}_{\nu)}+f_{ab}u^{a}_{\mu}u^{b}_{\nu}~~,
\eeq
\beq
f^{\perp}_{\mu a}={\perp^{\lambda}}_{\mu}\nabla_{a}f^{\perp}_{\lambda}+\left(\nabla_{a}u^{b}_{\mu}\right)\nabla_{b}f~~,~~f_{ab}=\nabla_{(a}\nabla_{b)}f-f_{\mu}^{\perp}\nabla_{b}u_{a}^{\mu}~~.
\eeq
On the world-volume the independent components are $f_{\mu\nu}^{\perp},f_{\mu}^{\perp},f$. Here the symbol $\perp$ means that the space-time indices are purely orthogonal to the world-volume, for example, ${u^{\mu}}_{a}f_{\mu}^{\perp}=0$. Using this decomposition we obtain the constraint equations
\beq
\perp^{\lambda}_{\nu}\perp^{\sigma}_{\rho}B^{\mu(\nu\rho)}k_{\mu}=0~~,
\eeq
\beq\label{c1}
\perp^{\sigma}_{\nu}B^{\mu\nu}k_{\mu}=\perp^{\sigma}_{\nu}B^{\nu\mu\rho}\nabla_{\mu}k_{\rho}+\perp^{\sigma}_{\nu}\nabla_{a}\left(B^{\nu\mu\lambda}u^{a}_{\nu}k_{\mu}+\perp^{\lambda}_{\rho}B^{\mu\rho\nu}u_{\nu}^{a}k_{\mu}\right)~~,
\eeq
which are trivially satisfied due to the equations of motion for \eqref{pdstress} obtained from solving $\nabla_{\nu}\hat{T}^{\mu\nu}=0$ , namely \cite{Vasilic:2007wp}
\beq
\perp^{\lambda}_{\nu}\perp^{\sigma}_{\rho}B^{\mu(\nu\rho)}=0~~,
\eeq
\beq\label{c11}
\perp^{\sigma}_{\nu}B^{\mu\nu}=\perp^{\sigma}_{\nu}\nabla_{a}\left(B^{\nu\mu\lambda}u^{a}_{\nu}+\perp^{\lambda}_{\rho}B^{\mu\rho\nu}u_{\nu}^{a}\right)~~.
\eeq
In fact, in order to obtain \eqref{c1} from \eqref{c11} it is only necessary to contract \eqref{c11} with $k_{\mu}$. Finally, we obtain the non-trivial current conservation equation 
\beq \label{eq1}
\begin{split}
\nabla_{a}\Big(B^{ab}k_{b}+u^{a}_{\mu}k_{\nu}^{\perp}\nabla_{c}\left(B^{\mu\rho\nu}u_{\rho}^{c}\right)&+B^{a\mu\rho}\nabla_{\rho}k_{\mu}+B^{\mu\nu\rho}k_{\mu}u_{\nu}^{b}{K^{a}}_{b\rho}~~ \\ \\
&+{u_{\mu}}^{a}k_{\nu}^{\perp}\nabla_{c}\left(\perp^{\nu}_{\lambda}B^{\mu\lambda c}\right)+B^{\mu\rho b}k_{\mu}{K^{a}}_{b\rho}-\nabla_{b}\left(B^{\mu(ab)}k_{\mu}\right)\Big)=0~~,
\end{split}
\eeq
where we have defined $k^{\perp}_\mu=\perp^{\lambda}_{\mu}k_{\lambda}$ and used Eq.~\eqref{c11}. The second line in \eqref{eq1} is in fact pure gauge as it involves only the components $B^{\mu\nu a}$, which can be set to zero everywhere on the world-volume, except at the boundary, using the covariance of \eqref{pdstress} \cite{Vasilic:2007wp}. We therefore use this freedom to set $B^{\mu\nu a}=0$ on the interior of the world-volume and hence find the conserved surface currents
\beq
\mathcal{P}_{k}^{a}=B^{ab}k_{b}+u^{a}_{\mu}k_{\nu}^{\perp}\nabla_{c}\left(B^{\mu\rho\nu}u_{\rho}^{c}\right)+B^{a\mu\rho}\nabla_{\rho}k_{\mu}+B^{\mu\nu\rho}k_{\mu}u_{\nu}^{b}{K^{a}}_{b\rho}~~.
\eeq
This set of conserved surface currents coincides with those identified in \cite{Armas:2013hsa} and presented in Eq.~\eqref{conc} provided one makes the identifications \cite{Armas:2013hsa}
\beq
B^{ab}=T^{ab}+2\mathcal{D}^{(aci}{K^{b)}}_{ci}~~,~~B^{abi}=-\mathcal{D}^{abi}~~,~~B^{aij}=\mathcal{S}^{aij}~~.
\eeq
The conservation equation \eqref{eq1} is invariant under field redefinitions which can be clearly seen using the transformations \eqref{extra2}-\eqref{extra21}. Furthermore, the current $\mathcal{P}_{k}^{a}$ is not invariant under these field redefinitions and in fact transforms as
\beq
\mathcal{P}_{k}^{a}\to\mathcal{P}_{k}^{a}-B^{ab}k_{b}u^{c}_{\rho}\nabla_{c}\tilde\varepsilon^{\rho}_{\perp}~~,
\eeq
where we have decomposed $\tilde\varepsilon^{\mu}$ as $\tilde\varepsilon^{\mu}={u^{\mu}}_{a}\tilde\varepsilon^{a}+\tilde\varepsilon^{\mu}_\perp$ and considered the only non-trivial variation along orthogonal directions to the world-volume. Nevertheless, the expressing for the charges obtained using $\mathcal{P}_{k}^{a}$ given by Eq.~\eqref{ccharge} are invariant under these transformations.


\subsection*{Conservation of surface currents at the boundary}
In the analysis above we have only considered the interior of the world-volume but solving \eqref{tosolve} also results in boundary conditions. On the boundary we need to decompose $\nabla_b f$ as $\nabla_b f=\eta_b\nabla_\perp f + v_{b}^{\tilde a}\nabla_{\tilde a}f$ where $\eta_b$ is a unit normal vector to the world-volume boundary, $v_{b}^{\tilde a}$ are projectors onto the world-volume boundary and the indices $\tilde a$ denote directions along the boundary \cite{Armas:2012ac, Armas:2013aka}. The independent components are thus $f_\mu^{\perp},\nabla_\perp f,f$. With this we find the two constraint equations
\beq
B^{\nu\mu\rho}k_{\mu}\perp^{\lambda}_{(\rho}{u_{\nu)}}^{a}\eta_a|_{\partial \mathcal{W}_{p+1}}=0 ~~,~~B^{\mu(\nu\rho)}k_{\mu}{u_{\nu}}^{a}{u_{\rho}}^{b}\eta_a\eta_b|_{\partial \mathcal{W}_{p+1}}=0~~,
\eeq
which are trivially satisfied due to the equations for the stress-energy tensor, which in turn are equal to the two above equations but without the contraction with $k_{\mu}$  \cite{Vasilic:2007wp}. Lastly, we find the conservation equation for the world-volume current
\beq
\nabla_{\tilde a}\left(B^{\mu(ab)}k_\mu v_{b}^{\tilde a}\eta_a\right)-\eta_a\mathcal{P}^{a}_{k}=0~~,
\eeq
where now we write the full current $\mathcal{P}^{a}_{k}$ including the components $B^{\mu\nu a}$, which cannot be set to zero at the boundary \cite{Vasilic:2007wp}, using \eqref{eq1}, that is
\beq 
\begin{split}
\mathcal{P}^{a}_{k}=B^{ab}k_{b}+u^{a}_{\mu}k_{\nu}^{\perp}\nabla_{c}\left(B^{\mu\rho\nu}u_{\rho}^{c}\right)&+B^{a\mu\rho}\nabla_{\rho}k_{\mu}+B^{\mu\nu\rho}k_{\mu}u_{\nu}^{b}{K^{a}}_{b\rho}~~ \\ \\
&+{u_{\mu}}^{a}k_{\nu}^{\perp}\nabla_{c}\left(\perp^{\nu}_{\lambda}B^{\mu\lambda c}\right)+B^{\mu\rho b}k_{\mu}{K^{a}}_{b\rho}-\nabla_{b}\left(B^{\mu(ab)}k_{\mu}\right)~~.
\end{split}
\eeq


\section{Spin current conservation from the action} \label{spin}
The action \eqref{fullaction} when varied with respect to the scalars $X^{\mu}(\sigma^a)$ yields the pole-dipole equations equations of motion upon suitable identifications \cite{Armas:2013hsa}. However, the general form of the pole-dipole equations \cite{Vasilic:2007wp} requires the spin current $\mathcal{S}^{a\mu\nu}$ to be conserved according to second equation in \eqref{spinconservation} as well as the integrability condition \cite{Armas:2013hsa}
\beq
\mathcal{D}^{ab[i}{K_{ab}}^{j]}=0~~.
\eeq
Here we will show that both these equations can be derived from the action \eqref{fullaction} with Lagrangian \eqref{lagrangian} by requiring it to be invariant under changes of the normal fields ${n^{\mu}}_{i}$. The extrinsic twist potential ${\omega_{a}}^{ij}$ is, in a sense, a gauge dependent object as it transforms under different choices of the normal fields which form a SO(2) gauge group for each transverse two-plane. In particular, we consider an infinitesimal transformation of the form
\beq \label{normal}
{n^{\mu}}_i\to{\tilde n^{\mu}}_i={n^{\mu}}_i+\nabla^{\mu}\Lambda_i~~.
\eeq
By definition, we must have that ${\tilde n^{\mu}}_i{\tilde n_{\mu j}}=\delta_{ij}$, implying, since ${n^{\mu}}_i{n_{\mu j}}=\delta_{ij}$ and ${u_{\mu}}^{a}{\tilde n^{\mu}}_i=0$ that
\beq
\nabla^{(i}\Lambda^{j)}=0~~,~~{u_{\mu}}^{a}\nabla^{\mu}\Lambda^{i}=0~~.
\eeq 
Both the extrinsic curvature and the extrinsic twist potential transform under \eqref{normal}. Using the definition ${K_{\mu\nu}}^{\rho}=-{\gamma^{\sigma}}_{\nu}{n^{\rho}}_{i}{\gamma^{\lambda}}_{\mu}\nabla_{\lambda}{n_{\sigma}}^{i}$ together with $\omega_a {}^{\lambda\rho} = {n^{\lambda}}_{i} \nabla_a n^{\rho i}$ we find
\beq
{u^{\mu}}_{a}{u^{\nu}}_{b}{n_\rho}^{i}\delta_{\Lambda}{K_{\mu\nu}}^{\rho}=-{u^{\mu}}_{b}\nabla_{a}\nabla_{\mu}\Lambda^{i}-{u^{\mu}}_{b}\nabla^{i}\Lambda_{j}\nabla_a{n_\mu}^{j}~~,
\eeq
\beq
{n_\lambda}^{i}{n_\rho}^{j}\delta_{\Lambda}{\omega_{a}}^{\lambda\rho}=-\nabla_\mu\Lambda^j \nabla_a {n^{\mu i}}-{n_{\mu}}^{j}\nabla_b\nabla^{\mu}\Lambda^{i}~~,
\eeq
where we have ignored terms of order $\mathcal{O}(\Lambda_i^2)$. The outer curvature of the embedding ${\Omega_{ab}}^{ij}$ defined as
\beq
{\Omega_{ab}}^{ij}=\nabla_{a}{\omega_{b}}^{ij}-\nabla_{b}{\omega_{a}}^{ij}+{\omega_{a}}^{ik}{\omega_{bk}}^{j}-{\omega_{b}}^{ik}{\omega_{ak}}^{j}~~,
\eeq
can be seen as a field strength for the field ${\omega_{a}}^{ij}$ and under \eqref{normal} transforms as
\beq
\Delta_\Lambda {\Omega_{ab}}^{ij}=2\nabla_{k}\Lambda^{[i}{\Omega_{ab}}^{j]k}~~.
\eeq
For the gauge transformation to preserve ${\Omega_{ab}}^{ij}$ one must impose $\nabla_{k}\Lambda^{[i}{\Omega_{ab}}^{j]k}=0$. Turning now to the action \eqref{fullaction} and ignoring the first four terms in the the Lagrangian \eqref{lagrangian} since they do not transform under \eqref{normal} we find the variation
\beq \label{lvar}
\begin{split}
\delta_\Lambda I&=\beta\int_{\mathcal{B}_p}dV_{(p)}R_0\left({\mathcal{D}^{\mu\nu}}_{\rho}\delta_\Lambda {K_{\mu\nu}}^{\rho}+{\mathcal{S}^{a}}_{\lambda\rho}\delta_\Lambda{\omega_{a}}^{\lambda\rho}\right)\\
&=-\beta\int_{\mathcal{B}_p}dV_{(p)}R_0\left[\left(2\mathcal{D}^{ab[i}{K_{ab}}^{j]}+{n_\rho}^{i}{n_{\lambda}}^{j}\nabla_{a}\mathcal{S}^{a\lambda\rho}\right)\nabla_{i}\Lambda_j+\nabla_{b}\left(\mathcal{S}^{bij}\nabla_{i}\Lambda_{j}\right)\right]~~.
\end{split}
\eeq
Since this variation must hold for all $\Lambda^{i}$ we have that 
\beq \label{integra}
2\mathcal{D}^{ab[i}{K_{ab}}^{j]}=-{n_\rho}^{i}{n_{\lambda}}^{j}\nabla_{a}\mathcal{S}^{a\lambda\rho}~~,
\eeq
and the last term in \eqref{lvar} yields a boundary term such that
\beq \label{boundary}
\mathcal{S}^{aij}\eta_a|_{\partial\mathcal{B}_p}=0~~,
\eeq
where $\eta_a$ is a unit normal vector to the boundary of the spatial world-volume $\mathcal{B}_p$. The conditions \eqref{integra} and \eqref{boundary} fit into the pole-dipole equations of motion obtained in form given in \cite{Armas:2013hsa} provided the l.h.s. of \eqref{integra} vanishes, which is indeed the case for the Lagrangians \eqref{lagrangian} we consider. Therefore the r.h.s. of \eqref{integra} also vanishes individually and hence we obtain the conservation equation given in \eqref{spinconservation}.

\section{Relations for derivatives of Killing vector fields} \label{trans_spin}

We compute in this appendix the quantity $n_i {}^\mu n_j {}^\nu \nabla_\mu k_{\nu} |_{\CB_p}$
 where $k^\mu$ is a Killing vector field of the background being either one of the Killing vector fields with non-trivial orbit on the brane embedding $\xi$ and $\chi_q$ or one of the transverse spin Killing vector fields $\hat{\chi}_{\hat{q}}$.  

Consider a transverse spin plane with Killing vector field $\hat{\chi}_{\hat{q}} = \partial_\phi$. The metric for such a plane can be written as
\begin{equation}
\delta_{ij} n^i {}_\mu n^j {}_\nu dx^\mu dx^\nu = H (dr^2 + r^2 d\phi^2 )~~,
\end{equation}
where $i,j$ only runs over the two values corresponding to the particular spin plane and $H$ does not depend on $\phi$. The normal vectors are
\begin{equation}
n^1 {}_r = \sqrt{H} \cos \phi \spa n^1 {}_\phi = - \sqrt{H }r \sin \phi \spa n^2 {}_r = \sqrt{H} \sin \phi \spa n^2 {}_\phi = \sqrt{H} r \cos \phi ~~.
\end{equation}
We compute for a generic Killing vector field $k^\mu$
\begin{eqnarray}
n_i {}^\mu n_j {}^\nu \nabla_\mu k_\nu &=& n_i {}^\mu n_j {}^\nu \nabla_{[\mu} k_{\nu]} = n_i {}^\mu n_j {}^\nu \partial_{[\mu} k_{\nu]} = \epsilon^{(\hat{q})}_{ij}
 \frac{1}{2} g^{rr} g^{\phi\phi} ( n_{1r} n_{2\phi} - n_{1\phi} n_{2r} ) \partial_r k_{\phi} \nn \\ &=& \epsilon^{(\hat{q})}_{ij}
 \frac{1}{2} g^{rr} g^{\phi\phi} ( n_{1r} n_{2\phi} - n_{1\phi} n_{2r} ) \partial_r (g_{\phi\mu} k^\mu) ~~,
\end{eqnarray}
where we used that $\partial_\phi k_r=0$ since $\hat{\chi}_{\hat{q}} =\partial_\phi$ is a Killing vector field. Consider first $k^\mu = \hat{\chi}_{\hat{q}}^\mu$. We compute
\begin{equation}
n_i {}^\mu n_j {}^\nu \nabla_\mu \hat{\chi}_{\hat{q},\nu} =  \epsilon^{(\hat{q})}_{ij}
\frac{1}{2} g^{rr} g^{\phi\phi} ( n_{1r} n_{2\phi} - n_{1\phi} n_{2r} ) \partial_r g_{\phi\phi} =\epsilon^{(\hat{q})}_{ij}
( 1 + \frac{1}{2} r \partial_r H)~~.
\end{equation}
Evaluating this on the brane (which sits at $r=0$ as explained in Section \ref{thermo}) we get
\begin{equation}
\label{imp_id}
n_i {}^\mu n_j {}^\nu \nabla_\mu \hat{\chi}_{\hat{q},\nu} \Big|_{\CB_p} = \epsilon^{(\hat{q})}_{ij}~~,
\end{equation}
since regularity at $r=0$ requires $\partial_r H |_{r=0} = 0$.

Consider instead $k^\mu = \xi^\mu$. We choose coordinates such that $\xi = \partial_t$.  Then,
\begin{equation}
n_i {}^\mu n_j {}^\nu \nabla_\mu \xi_{\nu} =  \epsilon^{(\hat{q})}_{ij}
\frac{1}{2} g^{rr} g^{\phi\phi} ( n_{1r} n_{2\phi} - n_{1\phi} n_{2r} ) \partial_r g_{\phi t} ~~.
\end{equation}
Instead one finds
\begin{eqnarray}
\xi^a \omega_{aij} &=& \omega_{t ij} = n_{i\mu} \nabla_t n_j {}^{\mu} = n_{i\mu} \Gamma^\mu_{t\sigma} n_j {}^{\sigma} = n_i {}^{\mu} \frac{1}{2} ( \partial_\sigma g_{t \mu} - \partial_\mu g_{t\sigma} ) n_j {}^{\sigma} \nn \\ &=& - \epsilon_{ij}^{(\hat{q})} \frac{1}{2} g^{rr} g^{\phi\phi} ( n_{1r} n_{2\phi} - n_{1\phi} n_{2r} ) \partial_r g_{t\phi}~~. 
\end{eqnarray}
Hence
\begin{equation}
\label{nablaxi}
n_i {}^\mu n_j {}^\nu \nabla_\mu \xi_{\nu} \Big|_{\CB_p} = - \xi^a \omega_{aij}
\spa 
n_i {}^\mu n_j {}^\nu \nabla_\mu \chi_{q,\nu} \Big|_{\CB_p} = - \chi_q^a \omega_{aij}~~,
\end{equation}
where the second identity can be obtained for $k^\mu = \chi_q^\mu$ in the same way as for $k^\mu=\xi^\mu$.


\section{Generating numerical charged black rings} \label{nbr}
In this appendix we take the black ring solution numerically constructed in \cite{Dias:2014cia} in $D=7$ and construct numerical charged black rings in Einstein-Maxwell-Dilaton theory via a solution generating technique. This solution generating technique consists in uplifting a vacuum solution of pure Einstein gravity by adding an extra flat direction $z$, consequently boosting it along the $(t,z)$ plane with boost parameter $\alpha$ and later reducing the metric over the $z$-direction. This uplift, boost and reduce (UBR) transformation yields a charged solution of the actions of motion derived by varying the action
\beq
S=\int\sqrt{-g}\left(R-\frac{1}{2}(\partial \Phi)^2-\frac{1}{4}e^{-(D-1)a\Phi}F^2_{[2]}\right)~~,~~a=\frac{1}{2(D-1)(D-2)}~~,
\eeq
where $R$ is the Ricci curvature tensor, $\Phi$ the dilaton field and $F_{[2]}$ the field strength for the gauge field $A_{[1]}$. 

The neutral black ring solutions constructed in \cite{Dias:2014cia} for $D=6,7$ are characterised by the following metric
\beq \label{neutralansatz}
ds^2=-Adt^2+Bdy^2+C(dx+Fdy)^2+S_1(d\psi-Wdt)^2+S_2d\Omega^2_{(D-4)}~~,
\eeq
where all the functions $A,B,C,F,S_1,W,S_2$ depend only on the coordinates $x$ and $y$. Applying the UBR transformation we obtain the transformed metric \footnote{See App. C of \cite{Armas:2013aka} for details on this transformation.}
\beq \label{chargedmetric}
\begin{split}
ds^2_{Q}=H^{\frac{1}{D-3}}\left(-\frac{A}{H}\left(dt^2+S_1\sinh^2\alpha d\psi^2\right)+\frac{S_1}{H}\left(\cosh\alpha d\psi -Wdt\right)^2+C(dx+Fdy)^2 +S_2d\Omega^2_{(D-4)}\right)~~,
\end{split}
\eeq
where the dilaton field $\Phi$ is defined via the relation 
\beq
e^{-2(D-2)a\Phi}\equiv H=1+\sinh^2\left(1-A+S_1W^2\right)~~,
\eeq
and the one-form gauge field in turn reads
\beq \label{chargedgauge}
A_\mu dx^\mu=\frac{\sinh\alpha}{H}\left(\cosh\alpha(1-A+S_1W^2)dt-S_1Wd\psi\right)~~.
\eeq

Defining $M^{*},J^*,\Omega^*,T^*$ as the thermodynamic quantities associated with the neutral solution \eqref{neutralansatz} before the UBR transformation, then the new quantities corresponding to the solution \eqref{chargedmetric}-\eqref{chargedgauge} are related to the neutral ones via the relations
\beq \label{newm}
M=\frac{M^*}{(D-2)}(1+(D-3)\cosh^2\alpha)~~,~~J=J^*\cosh\alpha~~,~~\Omega=\frac{\Omega^*}{\cosh\alpha}~~,~~T=\frac{T^*}{\cosh\alpha}~~.
\eeq
Furthermore, one finds that the new solution has total charge and chemical potential given by, respectively,
\beq\label{newq}
Q=\frac{(D-3)}{(D-2)}M^*\sinh\alpha\cosh\alpha~~,~~\Phi_H=\tanh\alpha~~.
\eeq
Introducing now the reduced quantities \eqref{reducedthermo} using the new thermodynamic quantities \eqref{newm}-\eqref{newq} and given the numerical data $j^{*},a_{\text{H}}^{*},\omega_{\text{H}}^{*},t_{\text{H}}^{*}$ corresponding to the neutral black ring solution of \cite{Dias:2014cia} we are able to generate numerical charged black rings in $D=7$ and plot their phase diagram in Sec.~\ref{sec:examples} for different values of $\Phi_H$.

\addcontentsline{toc}{section}{References}
\footnotesize
\providecommand{\href}[2]{#2}\begingroup\raggedright\endgroup

\end{document}